\documentclass[twocolumn,prd,showpacs,superscriptaddress]{revtex4}
\usepackage{graphicx}
\usepackage{bm}
\usepackage{amsfonts}
\usepackage[latin1]{inputenc} 
\usepackage[intlimits]{amsmath}
\usepackage{amsxtra}
\usepackage{amssymb}
\usepackage{psfrag}
\usepackage{amstext}
\newlength{\twocolumnwidth}
\newlength{\auxl}

\newlength{\auxlv}

\newcommand{\Eps}{Eps/} 
\renewcommand{\Eps}{} 
\newlength{\DL}

\newlength{\pwd} 
\begin{document} 
\title{Causal signal transmission by quantum fields.%
\\ 
IV: The causal Wick theorem.%
} 
\author{L.\ I.\ Plimak} 
\affiliation{Institut f\"ur Quantenphysik, Universit\"at Ulm, 
D-89069 Ulm, Germany.} 
\author{S.\ Stenholm} 
\affiliation{Institut f\"ur Quantenphysik, Universit\"at Ulm, 
D-89069 Ulm, Germany.} 
\affiliation{Physics Department, Royal Institute of Technology, KTH, Stockholm, Sweden.} 
\affiliation{Laboratory of Computational Engineering, HUT, Espoo, Finland.} 
\date{\today} 
\begin{abstract} 
Wick's theorem in the Schwinger-Perel-Keldysh closed-time-loop formalism is written in a form where the place of contractions is taken by the linear response function of the field. This result demonstrates that the physical information supplied by Wick's theorem for operators is propagation of the free field in space and time.
\end{abstract}
\maketitle 
\section{Introduction}
In the previous papers of these series \cite{API,APII,APIII} we investigated response properties of quantised fields. Analyses in those papers were, so to speak, pure kinematics. Starting from this paper we extend our analyses to dynamics. Here we derive the formal tool on which other dynamical results depend, and which we term the {\em causal Wick theorem\/}. The latter is a functional (Hori's) \cite{Hori,VasF} form of {\em Wick's theorem\/} \cite{Wick,Schweber,Bogol} in the Schwinger-Perel-Keldysh {\em closed-time-loop formalism\/} \cite{SchwingerC,Perel,Keldysh}, written in terms of the {\em linear response function\/} \cite{Kubo} characteristic of the field in question. The term {\em causal Wick theorem\/} was introduced in the unpublished paper \cite{Diag}. 

In the standard Feynman-Dyson technique of quantum field theory \cite{Schweber}, Wick's theorem supplies dynamical information about free fields, expressed by their {\em contractions\/}, or {\em Feynman propagators\/} (we avoid the term {\em causal Green function\/}, because we associate ``causal'' with ``retarded,'' as in {\em causal response\/} and {\em causal space-time evolution\/}). In the closed-time-loop formalism, propagators becomes matrices, but the fact that Wick's theorem is a supplier of dynamical information contained in propagators does not change. 

As was shown in \mbox{Refs.\ \cite{API,APIII}} (see also \cite{Corresp,BWO}), components of the Perel-Keldysh matrix propagator may be expressed by the linear response function (known in quantum field theory as the retarded Green function \cite{Schweber,Bogol}) characteristic of the field. As a physical quantity, this function emerges in Kubo's linear response theory \cite{Kubo} for the field in question, see \mbox{Refs.\ \cite{API,APIII}} and 
sections \ref{ch:TF} and \ref{ch:OR} of this paper.
Linear response of the free electromagnetic field was considered by Schwinger \cite{SchwingerR}. The only information about the field needed for Wick's theorem is its response. Thus the formal operation of operator reordering which is the subject of Wick's theorem is inherently connected to how the field propagates in space-time. One may say that propagation of the field is the actual physical content of Wick's theorem. The causal Wick theorem makes this evident. 

An immediate word of caution is in place here. In the literature, the term {\em Wick's theorem\/} is used for two different facts. Wick's original theorem expresses time-ordered products of free-field {\em operators\/} as linear combinations of normally-ordered products of the same operators. This theorem holds irrespective of the quantum state of the system and {\em ipso facto\/} may be used for systems out of thermal equilibrium. The other meaning of this term is a Gaussian factorisation of {\em Green functions\/} of free fields in thermal equilibrium (underlying, e.g., the Matsubara diagram techniques \cite{Matsubara}). Phase-space methods may be applied to quantum dynamics in both cases, cf.\ \mbox{Refs.\ \cite{BWO}} and \cite{Imag,TrapFerm}, respectively. This particular paper is part of the effort directed at extending the out-of-equilibrium methods of \mbox{Ref.\ \cite{BWO}} to space-time evolution of arbitrary interacting quantum fields, cf.\ the introduction to our paper \cite{API}. The question of how far the equilibrium phase-space techniques of \mbox{Refs.\ \cite{Imag,TrapFerm}} may be generalised is outside the scope of this paper (and very much outside the scope of this series of papers). Consequently, here the term {\em Wick's theorem\/} is used strictly in the sense of a relation between time-ordered and normally-ordered operator products. Had the need arisen to emphasise the said distinction, we shall talk about {\em Wick's theorem for operators\/}. 

This paper is self-contained and does not directly rely on our previous work. A brief overview of papers \cite{API,APII,APIII} seems nonetheless beneficial for putting this one in perspective. In \mbox{Refs.\ \cite{API,APII,APIII}} we demonstrated a structural equivalence of response properties of stochastic c-number and quantised fields. It applies without reservation to an arbitrary bosonic field, whether real or complex, free or interacting, in or out of equilibrium. At the cost of exploiting essentially nonclassical objects, such as Grassmann sources \cite{APIII,VasF,Beresin}, it extends also to fermionic fields. The technical tool making this equivalence evident is {\em response transformation\/} \cite{API,APII,APIII}, cf.\ also \cite{Corresp,BWO}. Formally, it is a change of variables (termed {\em response substitution\/}) in the generating functionals of quantum averages of products of {Heisenberg}\ field operators ordered in the closed-time-loop style. In \cite{BWO}, a similar substitution was termed a change to {\em causal variables\/}. In this paper we use both terms: response substitutions introduce causal variables. 

In \cite{API,APII,APIII}, causal variables were identified for the harmonic oscillator, and then postulated for interacting fields. Papers \cite{APII,APIII} were concerned with consistency of such generalisation. It was shown that formal quantum response structures, which emerge as a result of the response transformation, coincide with typical response structures characteristic of stochastic c-number fields. In particular, classical reality and causality properties were shown to hold, exactly and without reservation, for quantised fields. 

The result of this and forthcoming papers in a nutshell is that {\em formal\/} response structures identified in \cite{API,APII,APIII}, firstly, naturally extend to underlying quantum dynamics, and, secondly, under {\em macroscopic\/} (also called mesoscopic \cite{endMeso}) conditions turn into {\em physical\/} response structures. The causal Wick theorem allows one to rewrite Dyson's standard perturbative approach of quantum field theory directly in the response representation. Similarly to how response transformations reveal the kinematical response properties of quantised fields, the causal Wick theorem reveals the dynamical response properties hidden in quantum equations of motion. For examples see \mbox{Refs.\ \cite{Corresp,BWO,QDynResp}}. 

The causal Wick theorem may be seen as a step towards dispelling the myth about inconsistency of space-time evolution with relativistic quantum field theory. The reason why the Feynman-Dyson type of approaches do not allow one to describe propagation of quantum fields in space and time is purely technical: they do not operate with sufficient number of Green functions. This becomes obvious already if looking at pair Green functions. In the Feynman-Dyson techniques, there exists only one pair Green function per field --- the dressed Feynman propagator. That this is not enough is obvious from a classical analogy. For a classical stochactic field, we have two pair quantities: the linear response function and the average of a product of two fields. In the quantum response viewpoint, the corresponding quantities are Kubo's linear response function \cite{Kubo} and the time-normal average of a pair of operators \cite{KelleyKleiner,GlauberTN,APII}. One dressed Feynman propagator cannot be equivalent to two independent quantities. 

In the Schwinger-Perel-Keldysh closed-time-loop formalism \cite{SchwingerC,Perel,Keldysh}, the number of independent pair Green functions doubles. Consequently, this formalism and the response representation are connected by a one-to-one transformation. This is all that response tranformations are about. 
Neither these transformations nor the causal Wick theorem may be introduced in the Feynman-Dyson techniques. Both are strictly subject to the closed-time-loop framework. 

We stress once again that we are talking about arbitrary nonequilibrium systems where fluctiations and response are not generally related to each other. In vacuum the time-normal average vanishes, while in equilibrium it is related to response by Kubo's fluctuation-dissipation theorem. In both cases one enjoys the luxury of having only one independent pair Green function to work with. This underlies success of the standard Feynman-Dyson techniques, and of the so-called Retarded-Advanced formalism \cite{Therm1,Therm2}, concerned with field propagation in interacting thermal systems \cite{endRA}. Our results, we repeat, apply to arbitrary nonequilibrium systems, where the said luxury is unavailable. 

We make extensive use of the functional (Hori's) form of Wick's theorem \cite{Hori,VasF} (the Hori-Wick theorem, for short). Outside the quantum-field-theoretical community, Wick's theorem is often regarded as something high-minded, and its functional form as high-minded in the extreme. In fact, the Hori-Wick theorem is a major simplification. Firstly, it is an excellent mnemonic tool. Its algebraic rigidity leaves one only with the freedom of formulae for contractions. These are worked out applying Wick's theorem to pair products, which is next to trivial (cf.\ sections \ref{ch:OM} and \ref{ch:W}). Secondly, it reduces complicated calculations to straighforward algebra (cf.\ sections \ref{ch:OF} and \ref{ch:CT}). Last but not least, the Hori-Wick theorem greatly simplifies extention of the formal techniques to fermions. By making use of auxiliary Grassmann variables \cite{Beresin,VasF,APIII}, sign rules characteristic of Wick's theorem for fermions are observed automatically, and the algebra remains mostly as for bosons. 

The paper is structured as follows. We start from deriving the causal Wick theorem for the harmonic oscillator in section \ref{ch:O}. In section \ref{ch:H} we extend it to the real (Hermitian) field. In both cases, results of \mbox{Ref.\ \cite{API}} form a coherent background for analyses in this paper. Unlike the real field, formulae we need for complex fields are absent in \cite{API,APIII}. Filling this gap is the subject of section \ref{ch:T}, which is a sizeable fraction of the formal effort in the paper. The causal Wick theorem for complex bosonic and fermionic fields is derived in section \ref{ch:S}. In section \ref{ch:N} we outline simplifications characteristic of the nonrelativistic\ (resonance, rotating wave) approximation. Appendices take care of formal particulars brushed under the carpet in the main body of the paper. They also outline details of derivations making the paper self-contained. Of independent interest to the reader may be appendices \ref{ch:DGF} and \ref{ch:B} which deal with the mathematical background. In appendix \ref{ch:DGF} we discuss variational derivatives by Grassmann fields. It is a somewhat streamlined version of the approach of \mbox{Ref.\ \cite{APIII}}. As a crucial generalisation, it includes chain rules for variational Grassmann derivatives, which are shown to be identical to those for conventional variational derivatives. In appendix \ref{ch:B}, we show that Wick's theorem may always be written as a Hori-Wick theorem; in a sense, this statement is more general than Wick's theorem itself.

\section{The causal Wick theorem for the harmonic oscillator}%
\label{ch:O}
\subsection{Hori's form of Wick's theorem}%
\label{ch:OH}
To better orient the reader we start from a simple example. Consider an harmonic oscillator with the Hamiltonian in the {Schr\"odinger}\ picture, 
{\begin{align}{{
 \begin{aligned} 
\hat H_0 = \hbar \omega _0\hat a^{\dag} \hat a , 
\end{aligned}}}%
\label{eq:19UX} 
\end{align}}%
where $\hat a^{\dag},\hat a$ are the bosonic creation/annihilation pair, 
{\begin{align}{{
 \begin{aligned} 
\ensuremath{\big[
\hat a,\hat a^{\dag}
\big]} = 1 . 
\end{aligned}}}%
\label{eq:20UY} 
\end{align}}%
In what follows we use condensed notation, 
{\begin{align}{{
\begin{aligned} 
f g & = \int dt f(t) g(t), \\ 
f K g & = \int dt dt' f(t) K(t-t')g(t'), 
\end{aligned}}}%
\label{eq:53WK} 
\end{align}}%
where \mbox{$K(t-t')$} is a c-number kernel, and $f(t)$, $g(t)$ may be c-numbers, q-numbers, or functional derivatives. Except for c-numbers, the order of factors in (\ref{eq:53WK}) matters. Notation (\ref{eq:53WK}) saves space and makes formulae much more transparent. It also greatly simplifies extention to quantised fields, cf.\ section \ref{ch:H} below. 

\begin{figure}[t]
\includegraphics[width=7.5cm]{\Eps 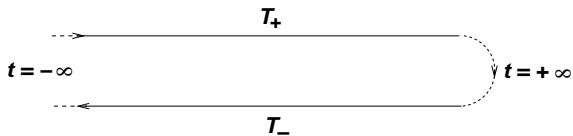}
\caption{The C-contour. }
\label{fig:C}
\end{figure}
We assume that the reader is familiar with the concepts of normal ordering, denoted \mbox{$
{\mbox{\rm\boldmath$:$}}
\cdots
{\mbox{\rm\boldmath$:$}} 
$}, and of the Schwinger-Perel-Keldysh's closed-time-loop, or C-contour, operator ordering \cite{SchwingerC,Perel,Keldysh}, denoted $T_C$. In brief, normal ordering puts all $\hat a^{\dag}$'s to the left of all $\hat a$'s. The $T_C$ ordering is best visualised as an ordering on the so-called C-contour depicted in fig.\ \ref{fig:C}. Operators positioned on the forward and reverse branches of the C-contour are marked by pluses and minuses, respectively. The ordering rule coincides with the travelling rule indicated in fig.\ \ref{fig:C}, so that operators on the forward (reverse) branches are time (reverse-time) ordered. For formal definitions see appendix \ref{ch:WO}. 

Now, let $\hat Q(t)$ be the time-dependent displacement operator, 
{\begin{align}{{
 \begin{aligned} 
\hat Q(t) & = \sqrt{\frac{\hbar }{2\omega _0}} \ensuremath{\big[
\hat a(t) + \hat a^{\dag}(t)
\big]} , & 
\hat a(t) & = \hat a\textrm{e}^{-i\omega _0 t}. 
\end{aligned}}}%
\label{eq:21UZ} 
\end{align}}%
Wick's theorem allows one to reorder $T_C$-ordered products of many $\hat Q$'s normally. Its functional (Hori's \cite{Hori,VasF}) form (the Hori-Wick theorem, for short) reads, 
{\begin{align}{{
 \begin{aligned} 
& {\begin{aligned}& T_C\mathcal{F}(\hat Q _+,\hat Q _-) = {\mbox{\rm\boldmath$:$}}
\mathcal{F}_N(\hat Q)
{\mbox{\rm\boldmath$:$}} , & 
\mathrm{where}\end{aligned}} 
\\ 
& \mathcal{F}_N( q_+) 
= \exp \mathcal{Z}_C\ensuremath{\bigg(
\frac{\delta }{\delta q_+},
\frac{\delta }{\delta q_-}
 \bigg)}
\mathcal{F}(q_+,q_-) \settoheight{\auxlv}{$|$}%
\raisebox{-0.3\auxlv}{$|_{q_-=q_+}$}. 
\end{aligned}}}%
\label{eq:59FF} 
\end{align}}%
In these relations, $\mathcal{F}(\cdot,\cdot)$ is an arbitrary c-number functional of two arguments, $q_{\pm}(t)$ are a pair of auxiliary c-number functional variables, and the {\em reordering form\/} \mbox{$
\mathcal{Z}_C(\cdot,\cdot)
$} is a functional bilinear form. In notation (\ref{eq:53WK}) it reads,
{\begin{multline}\hspace{0.4\columnwidth}\hspace{-0.4\twocolumnwidth} 
\mathcal{Z}_C\ensuremath{\big(
 \eta_+ ,
 \eta_- 
 \big)} \\ 
= -\frac{i\hbar}{2}
\ensuremath{\big[
\eta_+ G _{\text{F}} \eta_+ 
-\eta_- G _{\text{F}}^* \eta_- 
+ 2 \eta_- G^{(+)}\eta_+ 
\big]} , 
\hspace{0.4\columnwidth}\hspace{-0.4\twocolumnwidth} 
\label{eq:28VH} 
\end{multline}}%
where $ G _{\text{F}}$ and $ G^{(+)}$ are the Perel-Keldysh contractions, (cf.\ endnote \cite{endKern})
{\begin{align}{{
 \begin{aligned} 
& -i\hbar G_{\text{F}} (t-t') = \ensuremath{\big\langle 0\big|
T_+\hat Q (t)\hat Q (t')
\big|0\big\rangle} , 
\\
& -i\hbar G^{(+)} (t-t') = \ensuremath{\big\langle 0\big|
\hat Q(t)\hat Q(t')
\big|0\big\rangle} . 
\end{aligned}}}%
\label{eq:29VJ} 
\end{align}}%
The differentiations are performed using the formulae, 
{\begin{align}{{
 \begin{aligned} 
\frac{\delta q_+(t)}{\delta q_+(t')}
& = \frac{\delta q_-(t)}{\delta q_-(t')} 
=\delta(t-t'), \\
\frac{\delta q_+(t)}{\delta q_-(t')}
& = \frac{\delta q_-(t)}{\delta q_+(t')} 
=0 . 
\end{aligned}}}%
\label{eq:49WE} 
\end{align}}%
Derivation of the Hori-Wick theorem for the oscillator may be found, e.g., in our \mbox{Ref.\ \cite{API}}, appendix A. An alternative derivation may be constructed using the method of appendix \ref{ch:PW}, which is readily adapted to a real field. 
\subsection{``Mnemonic derivation'' of the Hori-Wick theorem}%
\begin{figure}[t]
\begin{center}
\includegraphics[width=170pt]{\Eps 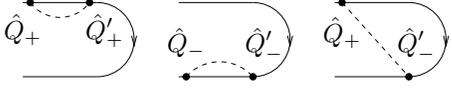}
\caption{The three possible ways a pair of displacement operators (bold dots) may be placed on the C-contour; $\hat Q=\hat Q(t)$, $\hat Q'=\hat Q(t')$. Dashed lines signify the corresponding contractions. }
\label{fig:contrR}
\end{center}
\end{figure}

\label{ch:OM}
The Hori-Wick theorem (\ref{eq:59FF}) calls for few remarks, which should benefit the reader without a strong quantum-field-theoretical background. As is shown in appendix \ref{ch:B}, an exponentiated bilinear form of functional derivatives dutifully generates ``all possible sets of contractions one can indicate, including the term without contractions'' \cite{Wick}, irrespective of whether the ``contractions'' make any physical sense. ``Contractions'' in this context are simply c-number kernels specifying the form. Hence {\em mere awareness\/} that Wick's theorem holds for a particular system allows one to immediatly write down the Hori-Wick theorem for it. It is convenient to firstly leave the reordering form unspecified, and then work it out applying the Hori-Wick theorem to pair products. For the oscillator, it means postulating eqs.\ (\ref{eq:59FF}), (\ref{eq:28VH}), with $ G _{\text{F}}$, $ G _{\text{F}}^*$ and $ G ^{(+)}$ regarded as a notation for three unknown kernels. They are found applying (\ref{eq:59FF}) to the products, 
{\begin{align}{{
 \begin{aligned} 
T_C \hat Q_+(t)\hat Q_+(t')& =T_+ \hat Q(t)\hat Q(t'), \\
T_C\hat Q_-(t)\hat Q_-(t')& =T_-\hat Q(t)\hat Q(t'), \\
T_C\hat Q_-(t)\hat Q_+(t')& =\hat Q(t)\hat Q(t'). 
\end{aligned}}}%
\label{eq:54WL} 
\end{align}}%
The corresponding $\mathcal{F}$'s are $q_+(t)q_+(t')$, $q_-(t)q_-(t')$ and $q_-(t)q_+(t')$. 
Applied to pair products, the reordering exponent in (\ref{eq:59FF}) may be reduced to two Taylor terms, 
{\begin{align}{{
 \begin{aligned} 
\exp \mathcal{Z}_C\ensuremath{\bigg(
\frac{\delta }{\delta q_+},
\frac{\delta }{\delta q_-}
 \bigg)} \to 1 + \mathcal{Z}_C\ensuremath{\bigg(
\frac{\delta }{\delta q_+},
\frac{\delta }{\delta q_-}
 \bigg)} . 
\end{aligned}}}%
\label{eq:57WP} 
\end{align}}%
It is easy to verify the formula, 
{\begin{align}{{
 \begin{aligned} 
\ensuremath{\bigg(
\frac{\delta }{\delta f}K\frac{\delta }{\delta g}
 \bigg)} f(t)g(t') = K(t-t'), 
\end{aligned}}}%
\label{eq:58WQ} 
\end{align}}%
where $K(t-t')$ is an arbitrary c-number kernel and $f(t),g(t)$ are two auxiliary c-number functions. Using it we obtain, 
{\begin{align}{{
 \begin{aligned} 
T_+ \hat Q(t)\hat Q(t') & = -i\hbar G _{\text{F}}(t-t') + {\mbox{\rm\boldmath$:$}}\hat Q(t)\hat Q(t'){\mbox{\rm\boldmath$:$}} , \\ 
T_- \hat Q(t)\hat Q(t') & = i\hbar G^* _{\text{F}}(t'-t) + {\mbox{\rm\boldmath$:$}}\hat Q(t)\hat Q(t'){\mbox{\rm\boldmath$:$}} , \\ 
\hat Q(t)\hat Q(t') & = -i\hbar G ^{(+)}(t-t') + {\mbox{\rm\boldmath$:$}}\hat Q(t)\hat Q(t'){\mbox{\rm\boldmath$:$}} . 
\end{aligned}}}%
\label{eq:55WM} 
\end{align}}%
Definitions (\ref{eq:29VJ}) follow as a consequence of the relation, 
{\begin{align}{{
 \begin{aligned} 
\ensuremath{\big\langle 0\big|
{\mbox{\rm\boldmath$:$}}\hat Q(t)\hat Q(t'){\mbox{\rm\boldmath$:$}} 
\big|0\big\rangle} = 0 . 
\end{aligned}}}%
\label{eq:56WN} 
\end{align}}%
Noteworthy is that three products (\ref{eq:54WL}) correspond to three possible ways of placing a pair of displacement operators on the C-contour (cf.\ fig.\ \ref{fig:contrR}), with in turn correspond to three unknown kernels specifying the form. Such ``mnemonic derivations'' allow one to quickly work out the details also in more complicated cases. 

\subsection{The test-case formula}%
\label{ch:OF}
The key formal results of \cite{API} may be recovered by considering the characteristic functional of vacuum averages of $T_C$-ordered products of $\hat Q(t)$, 
{\begin{align}{{
 \begin{aligned} 
\Phi_{\mathrm{vac}}(\eta_+,\eta_-) 
= \ensuremath{\big\langle 0\big| 
T_C\exp
\ensuremath{\big(
i\eta_+\hat Q_+ - i\eta_- \hat Q_-
 \big)} 
\big|0\big\rangle} . 
\end{aligned}}}%
\label{eq:26VE} 
\end{align}}%
We use notation (\ref{eq:53WK}). This quantity expresses vacuum fluctuations of the oscillator. It may be explicitly calculated using the Hori-Wick theorem (\ref{eq:59FF}), 
{\begin{multline}\hspace{0.4\columnwidth}\hspace{-0.4\twocolumnwidth} 
\Phi _{\textrm{vac}}\ensuremath{\big(
\eta _+,\eta _-
 \big)} 
\\ 
= \exp \mathcal{Z}_C\ensuremath{\bigg(
\frac{\delta }{\delta q_+},
\frac{\delta }{\delta q_-}
 \bigg)}
\exp\ensuremath{\big(
i\eta _+ q_+-i\eta _- q_-
 \big)}
\settoheight{\auxlv}{$|$}%
\raisebox{-0.3\auxlv}{$|_{ q_{\pm}=0}$} \\ 
= \exp \mathcal{Z}_C\ensuremath{\big(
i\eta_+ ,
-i\eta_- \big)}
, 
\hspace{0.4\columnwidth}\hspace{-0.4\twocolumnwidth} 
\label{eq:27VF} 
\end{multline}}%
cf.\ eq.\ (37) in \cite{API} (heed should also be payed to endnotes \cite{endPh,endKern}). This relation as well as similar formulae below will be called {\em test-case formulae\/}. The intermediate expression in (\ref{eq:27VF}) follows from (\ref{eq:59FF}), and from the fact that, for any $\mathcal{F}_N$, 
{\begin{align}{{
 \begin{aligned} 
\ensuremath{\big\langle 0\big| 
{\mbox{\rm\boldmath$:$}}
\mathcal{F}_N(\hat Q)
{\mbox{\rm\boldmath$:$}} 
\big|0\big\rangle} = \mathcal{F}_N(0). 
\end{aligned}}}%
\label{eq:52WJ} 
\end{align}}%
A proof of the final one may be constructed using the relation \cite{Corresp}, 
{\begin{align}{{
 \begin{aligned} 
\mathcal{F}_1\ensuremath{\bigg(
\frac{\delta }{\delta g}
 \bigg)} \mathcal{F}_2(g)\settoheight{\auxlv}{$|$}%
\raisebox{-0.3\auxlv}{$|_{g=0}$} = 
\mathcal{F}_2\ensuremath{\bigg(
\frac{\delta }{\delta f }
 \bigg)} \mathcal{F}_1(f )\settoheight{\auxlv}{$|$}%
\raisebox{-0.3\auxlv}{$|_{f =0}$} , 
\end{aligned}}}%
\label{eq:56TS} 
\end{align}}%
where $f(t)$, $g(t)$ are c-number functions and $\mathcal{F}_1(f )$, $\mathcal{F}_2(g)$ are c-number functionals. It can be verified expanding $\mathcal{F}_{1,2}$ in functional Taylor series. Applying it to (\ref{eq:55TR}) we find, 
{\begin{multline}\hspace{0.4\columnwidth}\hspace{-0.4\twocolumnwidth} 
\Phi _{\textrm{vac}}\ensuremath{\big(
\eta_+,\eta_-
 \big)} 
= 
\exp\ensuremath{\bigg(
i\eta _+\frac{\delta }{\delta \eta _+'}\bigg)} 
\exp\ensuremath{\bigg(
-i\eta _-\frac{\delta }{\delta \eta _-'}
 \bigg)}
\\ \times 
\exp \mathcal{Z}_C\ensuremath{\big(
\eta _+' ,
\eta _-'
 \big)}
\settoheight{\auxlv}{$|$}%
\raisebox{-0.3\auxlv}{$|_{\eta _{\pm}'=0}$} , 
\hspace{0.4\columnwidth}\hspace{-0.4\twocolumnwidth} 
\label{eq:57TT} 
\end{multline}}%
where $\eta _{\pm}'(t)$ are auxiliary c-number functions. 
The exponents with derivatives here are nothing but functional shift operators. Applying the shifts and setting $\eta _{\pm}'(t)$ to zero we recover eq.\ (\ref{eq:27VF}). 

\subsection{Response transformation revisited}%
\label{ch:OR}
{\em Response transformation\/} \cite{API} relates contractions (\ref{eq:29VJ}) to the retarded Green function of the oscillator, (cf.\ endnote \cite{endKern})
{\begin{align}{{
 \begin{aligned} 
 G _{\text{F}}(t-t') & = G _{\text{R}}^{(+)}(t-t') 
+ G _{\text{R}}^{(+)}(t'-t), \\ 
 G^{(+)}(t-t') & = G _{\text{R}}^{(+)}(t-t') 
- G _{\text{R}}^{(-)}(t'-t), 
\end{aligned}}}%
\label{eq:30VK} 
\end{align}}%
where $^{(\pm)}$ stands for separation of the frequency-positive and negative\ parts of a function (cf.\ appendix \ref{ch:RTF}). The kernel 
$G_{\text{R}}$ is Kubo's linear response function, 
{\begin{multline}\hspace{0.4\columnwidth}\hspace{-0.4\twocolumnwidth} 
G_{\text{R}}(t-t') = \frac{\delta \ensuremath{\big\langle 
{\hat{\mathcal Q}}(t)
\big\rangle} }{\delta J_{\mathrm{e}}(t')}\settoheight{\auxlv}{$\big|$}%
\raisebox{-0.3\auxlv}{$\big|_{J_{\mathrm{e}}=0}$} \\ 
= \frac{i}{\hbar }\theta(t-t') \ensuremath{\big[
\hat Q(t), \hat Q(t') 
\big]} , 
\hspace{0.4\columnwidth}\hspace{-0.4\twocolumnwidth} 
\label{eq:64WW} 
\end{multline}}%
defined with respect to the interaction 
with a c-number source $J_{\mathrm{e}}(t)$, 
{\begin{align}{{
 \begin{aligned} 
\hat H_{\text{I}}(t) = - \hat Q(t)J_{\mathrm{e}}(t)
\end{aligned}}}%
\label{eq:65WX} 
\end{align}}%
with ${\hat{\mathcal Q}}(t)$ being the Heisenberg displacement operator according to the Hamiltonian 
{\begin{align}{{
 \begin{aligned} 
\hat H(t) = \hat H_0 + \hat H_{\text{I}}(t) . 
\end{aligned}}}%
\label{eq:66WY} 
\end{align}}%
Explicitly, 
{\begin{align}{{
 \begin{aligned} 
 G _{\text{R}}(t-t')
= \theta(t-t')\frac{\sin \omega _0(t-t')}{\omega _0}
. 
\end{aligned}}}%
\label{eq:24VC} 
\end{align}}%
We stress that we introduce one retarded transfer (Green) function, rather than a retarded and an advanced Green functions, as, e.g., in 
\mbox{Ref.\ \cite{Therm1}}. 

Equations (\ref{eq:30VK}) suggest new variables \mbox{$
\eta(x,t),j_{\textrm{e}}(x,t)
$} ({\em causal variables\/}, in terminology of \cite{BWO}), such that, in notation (\ref{eq:53WK}), 
{\begin{align}{{
 \begin{aligned} 
\mathcal{Z}_C\ensuremath{\big(
 i\eta_+ ,
 -i\eta_- 
 \big)} = i\eta
 G _{\text{R}} j_{\textrm{e}} . 
\end{aligned}}}%
\label{eq:31VL} 
\end{align}}%
The corresponding {\em response substitution\/} is \cite{API}, 
{\begin{align}{{
 \begin{aligned} 
& \eta_{\pm} = \frac{ j_{\textrm{e}}}{\hbar }\pm \eta ^{(\mp)}, 
\end{aligned}}}%
\label{eq:33VN} 
\end{align}}%
with the inverse substitution being, 
{\begin{align}{{
 \begin{aligned} 
& \eta = 
\eta_+-\eta_-
, & 
 j_{\textrm{e}} = \hbar \ensuremath{\big[
\eta ^{(+)}_+ + \eta ^{(-)}_-
\big]} . 
\end{aligned}}}%
\label{eq:32VM} 
\end{align}}%
By omitting the arguments we will be able to reuse these relations literally for a real bosonic field. We thus obtain vacuum fluctuations in an astonishingly classical form, 
{\begin{align}{{
 \begin{aligned} 
\Phi _{\mathrm{vac}}(\eta_+,\eta_-)\settoheight{\auxlv}{$|$}%
\raisebox{-0.3\auxlv}{$|_{\mathrm{c.v.}}$} 
= \exp\ensuremath{\big(i\eta
 G _{\text{R}} j_{\textrm{e}}
 \big)} , 
\end{aligned}}}%
\label{eq:34VP} 
\end{align}}%
where c.v.\ (short for causal variables) refers to substitution (\ref{eq:33VN}). 

With (\ref{eq:33VN}) being an integral transformation that smears $\eta(x,t)$ all over the time axis, restoration of classical causality in (\ref{eq:34VP}) emerges as something of a miracle. Physical implications of a mere possibility of transformation (\ref{eq:33VN}) were discussed at length in Refs.\ \cite{Corresp,API,APII,APIII}. 

\subsection{Response transformation of the test-case formula}%
\label{ch:OT}
So far our discussion stays within the results of \mbox{Ref.\ \cite{API}}. The step we undertake in this paper is to extend response transformations to Wick's theorem, which in turn allows one to extend them further to nonlinear quantum dynamics. 

Formally, we want another pair of causal variables, $\zeta (t),q_{\mathrm{e}}(t)$, to replace $q_{\pm}(t)$, such that eq.\ (\ref{eq:27VF}) in new variables would read, 
{\begin{multline}\hspace{0.4\columnwidth}\hspace{-0.4\twocolumnwidth} 
\Phi _{\textrm{vac}}\ensuremath{\big(
\eta _+,\eta _-
 \big)}\settoheight{\auxlv}{$|$}%
\raisebox{-0.3\auxlv}{$|_{\mathrm{c.v.}}$} 
\\ 
= \exp \ensuremath{\bigg(
-i
\frac{\delta }{\delta q_{\textrm{e}}} 
 G _{\text{R}}
\frac{\delta }{\delta \zeta}
 \bigg)} 
\exp \ensuremath{\big(
i\eta q_{\textrm{e}}+i\zeta j_{\textrm{e}}
 \big)} 
\settoheight{\auxlv}{$|$}%
\raisebox{-0.3\auxlv}{$|_{ \zeta = q_{\mathrm{e}} =0}$} \\ 
= \exp \ensuremath{\big(
i\eta G _{\text{R}} j_{\textrm{e}}
 \big)} 
. 
\hspace{0.4\columnwidth}\hspace{-0.4\twocolumnwidth} 
\label{eq:35VQ} 
\end{multline}}%
We continue using notation (\ref{eq:53WK}). These variables may be identified comparing the linear forms in (\ref{eq:27VF}) and in (\ref{eq:35VQ}), 
{\begin{align}{{
 \begin{aligned} 
\eta _+ q_+-\eta _- q_-
= \eta q_{\textrm{e}}+\zeta j_{\textrm{e}}
. 
\end{aligned}}}%
\label{eq:36VR} 
\end{align}}%
Using (\ref{eq:33VN}) and the properties of the $^{(\pm)}$ operation (cf.\ eq.\ (\ref{eq:23KC}) in appendix \ref{ch:RTF}) we find, 
{\begin{multline}\hspace{0.4\columnwidth}\hspace{-0.4\twocolumnwidth} 
\eta _+ q_+-\eta _- q_- = 
\ensuremath{\bigg[
\frac{j_{\mathrm{e}}}{\hbar }+\eta^{(-)}
\bigg]} q_+
- 
\ensuremath{\bigg[
\frac{j_{\mathrm{e}}}{\hbar }-\eta^{(+)}
\bigg]} q_- \\ 
= 
j_{\mathrm{e}}\frac{q_+-q_-}{\hbar } 
+ 
\eta \ensuremath{\big[
q_+^{(+)}+q_-^{(-)}
\big]} . 
\hspace{0.4\columnwidth}\hspace{-0.4\twocolumnwidth} 
\label{eq:37VS} 
\end{multline}}%
This is satisfied by, 
{\begin{align}{{
 \begin{aligned} 
& \zeta = \frac{ q_+- q_-}{\hbar }, 
\ \ 
 q_{\textrm{e}} = q_+^{(+)} + q_-^{(-)} , 
\\ 
& q_{\pm} = q_{\textrm{e}}\pm\hbar \zeta^{(\mp)}. 
\end{aligned}}}%
\label{eq:38VT} 
\end{align}}%
The new variables assure correct transformation of the functional differential form, 
{\begin{align}{{
 \begin{aligned} 
\mathcal{Z}_C\ensuremath{\bigg(
\frac{\delta }{\delta q_+},
\frac{\delta }{\delta q_-}
 \bigg)} = -i
\frac{\delta }{\delta q_{\textrm{e}}} 
 G _{\text{R}}
\frac{\delta }{\delta \zeta}
. 
\end{aligned}}}%
\label{eq:39VU} 
\end{align}}%
Indeed, we know from \mbox{Ref.\ \cite{API}} that eq.\ (\ref{eq:31VL}) is a consequence of transformation (\ref{eq:30VK}) and substitution (\ref{eq:33VN}). Now note that eqs.\ (\ref{eq:31VL}) and (\ref{eq:39VU}) coincide up to the replacements, 
{\begin{align}{{
 \begin{aligned} 
& \eta_{\pm} \to \mp i\, \frac{\delta }{\delta q_{\pm}}, & 
& j_{\textrm{e}} \to -i\frac{\delta }{\delta \zeta}, & & \eta \to 
-i\frac{\delta }{\delta q_{\textrm{e}}} . 
\end{aligned}}}%
\label{eq:35KR} 
\end{align}}%
These replacements transform substitution (\ref{eq:33VN}) into, 
{\begin{align}{{
 \begin{aligned} 
\mp i\, \frac{\delta }{\delta q_{\pm}} = 
-\frac{i}{\hbar }\frac{\delta }{\delta \zeta} \mp 
i\ensuremath{\bigg(
\frac{\delta }{\delta q_{\textrm{e}}} 
 \bigg)}^{(\mp)}. 
\end{aligned}}}%
\label{eq:34KQ} 
\end{align}}%
Hence eq.\ (\ref{eq:39VU}) is found as a consequence of transformation (\ref{eq:30VK}) and substitution (\ref{eq:34KQ}). To show eq.\ (\ref{eq:39VU}) it suffices to verify eqs.\ (\ref{eq:34KQ}). 

By the standard chain laws, 
{\begin{align}{{
 \begin{aligned} 
\frac{\delta }{\delta q _{\pm}(t)} = \int dt' \ensuremath{\bigg[
\frac{\delta \zeta (t')}{\delta q _{\pm}(t)} 
\frac{\delta }{\delta \zeta (t')} 
+ 
\frac{\delta q _{\textrm{e}}(t')}{\delta q _{\pm}(t)}
\frac{\delta }{\delta q _{\textrm{e}}(t')}
\bigg]} . 
\end{aligned}}}%
\label{eq:78UR} 
\end{align}}%
Using (\ref{eq:38VT}) we obtain, 
{\begin{align}{{
 \begin{aligned} 
& \frac{\delta \zeta (t')}{\delta q _{\pm}(t)} 
= \pm\frac{1}{\hbar } \delta (t-t'), & 
& \frac{\delta q_{\textrm{e}}(t')}{\delta q _{\pm}(t)} 
= \delta^{(\pm)} (t'-t) , 
\end{aligned}}}%
\label{eq:79US} 
\end{align}}%
where $\delta^{(\pm)}(t)$ are the frequency-positive and negative\ parts of the delta-function (given by eq.\ (\ref{eq:40KW}) in appendix \ref{ch:RTF}). 
The second formula here is found writing the expression for $ q _{\textrm{e}}(t)$ explicitly as an integral transformation, 
{\begin{multline}\hspace{0.4\columnwidth}\hspace{-0.4\twocolumnwidth} 
 q_{\textrm{e}}(t) = \int dt'\ensuremath{\big[
\delta ^{(+)}(t-t') q_+(t') 
\\ 
+ \delta ^{(-)}(t-t') q_-(t')
\big]} , 
\hspace{0.4\columnwidth}\hspace{-0.4\twocolumnwidth} 
\label{eq:80UT} 
\end{multline}}%
cf.\ eqs.\ (\ref{eq:39KV}), (\ref{eq:40KW}) in appendix \ref{ch:RTF}. Combining (\ref{eq:78UR}) and (\ref{eq:79US}) and again recalling eqs.\ (\ref{eq:39KV}), (\ref{eq:40KW}) we have, 
{\begin{multline}\hspace{0.4\columnwidth}\hspace{-0.4\twocolumnwidth} 
\frac{\delta }{\delta q _{\pm}(t)} 
= \pm\frac{1}{\hbar } 
\frac{\delta }{\delta \zeta (t)} 
+ \int dt' \delta^{(\pm)} (t'-t)
\frac{\delta }{\delta q _{\textrm{e}}(t')} 
\\ 
= \pm\frac{1}{\hbar } 
\frac{\delta }{\delta \zeta (t)} 
+ \ensuremath{\bigg[
\frac{\delta }{\delta q _{\textrm{e}}(t)}
\bigg]}^{(\mp)} . 
\hspace{0.4\columnwidth}\hspace{-0.4\twocolumnwidth} 
\label{eq:81UU} 
\end{multline}}%
This coincides with (\ref{eq:34KQ}) up to overall phase factors. 

Few words regarding the choice of factors in eqs.\ (\ref{eq:31VL})--(\ref{eq:39VU}). To start with, Planck's constant is absent from the RHSs of eqs.\ (\ref{eq:31VL}) and (\ref{eq:35VQ}). The way $\hbar $ is distributed among the variables in (\ref{eq:38VT}) warrants that $q_{\mathrm{e}}$ has the dimension of the ``field'' $\hat Q$. Similarly, $j _{\mathrm{e}}(x,t)$ has the dimension of field source. The explicit $i$ factor in (\ref{eq:34VP}) assures consistency with conventional definitions of characteristic functions and functionals in probability theory and theory of random processes. When comparing results of this paper to \mbox{Ref.\ \cite{API}}, heed should be payed to endnotes \cite{endPh,endKern}. 
\subsection{The causal Wick theorem for operators}%
\label{ch:OO}
Transformation (\ref{eq:38VT}) was derived for vacuum averages, but nothing prevents one from applying it directly to eq.\ (\ref{eq:59FF}). As may be seen from eq.\ (\ref{eq:38VT}), the condition $q_+(t)=q_-(t)$ in terms of the causal variables becomes, 
{\begin{align}{{
 \begin{aligned} 
q_+(t)& =q_-(t)=q_{\mathrm{e}}(t), & \zeta (t) & = 0 . 
\end{aligned}}}%
\label{eq:44VZ} 
\end{align}}%
With this observation, and accounting for (\ref{eq:39VU}), we find the second of eqs.\ (\ref{eq:59FF}) in the form, 
{\begin{multline}\hspace{0.4\columnwidth}\hspace{-0.4\twocolumnwidth} 
 \mathcal{F}_N(q_{\mathrm{e}}) \\ 
=
\exp\ensuremath{\bigg(
-i
\frac{\delta }{\delta q_{\textrm{e}}} 
 G _{\text{R}}
\frac{\delta }{\delta \zeta}
 \bigg)} 
\mathcal{F}\ensuremath{\big(
q_{\mathrm{e}}+\hbar \zeta ^{(-)},q_{\mathrm{e}}-\hbar \zeta ^{(+)}
 \big)}\settoheight{\auxlv}{$|$}%
\raisebox{-0.3\auxlv}{$|_{\zeta =0}$} . 
\hspace{0.4\columnwidth}\hspace{-0.4\twocolumnwidth} 
\label{eq:48WD} 
\end{multline}}%
The first of eqs.\ (\ref{eq:59FF}) remains unchanged. Unlike (\ref{eq:35VQ}), eq.\ (\ref{eq:48WD}) holds irrespective of the quantum state of the system. 

\begin{center}* * *\end{center}

Equation (\ref{eq:48WD}) and its couterparts derived below are the main result of this paper. In what follows, eq.\ (\ref{eq:48WD}) will be extended to quantum fields in the true meaning of the word including fermions. 
\section{Extension to the real (Hermitian) field}%
\label{ch:H}
\subsection{Definitions and recollections}%
\label{ch:HD}
Since response transformations apply mode-wise, the case of the real field is a trivial generalisation of that of the harmonic oscillator. The free-field operator is, 
{\begin{align}{{
 \begin{aligned} 
\hat Q (x,t)
= \sum_{\kappa}\sqrt{\frac{\hbar}{2\omega_{\kappa }}} 
\ensuremath{\big[
u_{\kappa}(x) \hat a_{\kappa} \textrm{e}^{-i\omega _{\kappa}t } + 
u_{\kappa}^*(x) \hat a_{\kappa}^{\dag} \textrm{e}^{i\omega _{\kappa}t }
\big]} , 
\end{aligned}}}%
\label{eq:82UV} 
\end{align}}%
where \mbox{$\hat a_{\kappa }^{\dag},\hat a_{\kappa }$} are bosonic creation-annihilation pairs,
{\begin{align}{{
 \begin{aligned} 
{[\hat a_{\kappa},\hat a_{\kappa'}^{\dag}]} = \delta _{\kappa\kappa'}. 
\end{aligned}}}%
\label{eq:94VJ} 
\end{align}}%
The mode frequencies $\omega _{\kappa }$ and the c-number mode functions $u_{\kappa}(x)$ may be arbitrary. The meaning of the ``coordinate'' $x$ and of the ``mode index'' $\kappa $, as well as of the ``mode sum'' $\sum_{\kappa}$ and of the ``spatial integration'' $\int dx$, is problem-specific. Nothing precludes $\kappa $ from being (partly) continuous, and $x$ from being (partly) discrete. 

The case of a free bosonic field was amply covered in our \mbox{Ref.\ \cite{API}}. Transfer function of the real field is given by Kubo's formula, 
{\begin{multline}\hspace{0.4\columnwidth}\hspace{-0.4\twocolumnwidth} 
 G _{\text{R}}(x,x',t-t') 
= \frac{i}{\hbar }
\theta(t-t')\ensuremath{\big\langle 
\ensuremath{\big[
\hat Q (x,t),\hat Q (x',t')
\big]}
\big\rangle} . 
\hspace{0.4\columnwidth}\hspace{-0.4\twocolumnwidth} 
\label{eq:84UX} 
\end{multline}}%
The Perel-Keldysh contractions read, 
{\begin{align}{{
 \begin{aligned} 
& -i\hbar G_{\text{F}} (x,x',t-t') = \ensuremath{\big\langle 0\big|
T_+\hat Q (x,t)\hat Q (x',t')
\big|0\big\rangle} , 
\\
& -i\hbar G^{(+)} (x,x',t-t') = \ensuremath{\big\langle 0\big|
\hat Q (x,t)\hat Q (x',t')
\big|0\big\rangle} . 
\end{aligned}}}%
\label{eq:74AF} 
\end{align}}%
Explicit expressions for the kernels are found using (\ref{eq:82UV}). Relations for the contractions in terms of the transfer function were worked out 
in \mbox{Ref.\ \cite{API}}, 
{\begin{align}{{
 \begin{aligned} 
 G _{\text{F}}(x,x',t-t') & = G _{\text{R}}^{(+)}(x,x',t-t') 
\\ & \ \ \ + G _{\text{R}}^{(+)}(x',x,t'-t), \\ 
 G^{(+)}(x,x',t-t') & = G _{\text{R}}^{(+)}(x,x',t-t') 
\\ & \ \ \ - G _{\text{R}}^{(-)}(x',x,t'-t). 
\end{aligned}}}%
\label{eq:11JQ} 
\end{align}}%
Equations (\ref{eq:84UX}), (\ref{eq:74AF}) and (\ref{eq:11JQ}) generalise, respectively, eqs.\ (\ref{eq:24VC}), (\ref{eq:29VJ}) and (\ref{eq:30VK}). 
\subsection{The causal Wick theorem for the real field}%
\label{ch:HW}
Similarly to the case of the oscillator we make use of the condensed notation, 
{\begin{align}{{
 \begin{aligned} 
f g & = \int dx dt f(x,t) g(x,t), \\ 
f K g & = \int dx dx' dt dt' f(x,t) K(x,x',t-t')g(x',t'), 
\end{aligned}}}%
\label{eq:60TW} 
\end{align}}%
where \mbox{$K(x,x',t-t')$} is a c-number kernel, and $f(x,t)$, $g(x,t)$ may be q-numbers, c-numbers or variational derivatives. This notation applies throughout the rest of the paper. For the purposes of section \ref{ch:T} below, $f(x,t)$ and $g(x,t)$ may also signify Grassmann fields and Grassmann derivatives (see section \ref{ch:GV} and appendix \ref{ch:DGF}). Except for c-numbers, the order of factors in (\ref{eq:60TW}) matters. 

Understood in line with (\ref{eq:60TW}), all critical relations for the harmonic oscillator apply literally to the Hermitian field. This implies that eqs.\ (\ref{eq:24VC}), (\ref{eq:29VJ}) and (\ref{eq:30VK}) are substituted by (\ref{eq:84UX}), (\ref{eq:74AF}) and (\ref{eq:11JQ}), and that all functional variables 
acquire the second (spatial) argument, $\eta _{\pm}(t)\to\eta _{\pm}(x,t)$, $q _{\pm}(t)\to q_{\pm}(x,t)$, etc. Since the $^{(\pm)}$ operations apply only to the time variable, the spatial arguments are carried through all manipulations without affecting them. In particular, this applies to the transformation of the linear form (\ref{eq:36VR}) on which the derivation hinges. Response substitutions (\ref{eq:33VN}) and (\ref{eq:38VT}), which we have foresightedly written without arguments, apply literally. 

Here we briefly reiterate the logic of the argument adapted to the real-field case. 
\begin{itemize}
\item[1.]
Formulate the Hori-Wick theorem, eq.\ (\ref{eq:59FF}). This implies the definition of the reordering form by eq.\ (\ref{eq:28VH}), subject to notation (\ref{eq:60TW}), and of the contractions by eq.\ (\ref{eq:74AF}). 
\item[2.]
Use the Hori-Wick theorem to derive the test-case formula (\ref{eq:27VF}). 
\item[3.]
Apply response transformations to the final expression in the test-case formula, cf.\ eq.\ (\ref{eq:34VP}). 
\item[4.]
Work out the change of variables in the reordering form postulating transformation of the linear form (\ref{eq:36VR}). 
Verify transformation of the reordering form (\ref{eq:39VU}) by proving eqs.\ (\ref{eq:34KQ}). 
\item[5.]
Extend the causal Wick theorem to operators resulting in eq.\ (\ref{eq:48WD}). 
\end{itemize}
Steps 1--3 of this schedule are a result of \mbox{Refs.\ \cite{API,APIII}}, while steps 4 and 5 are the subject of this paper. In what follows this logic is further extended to complex fields, bosonic as well as fermionic. 

\section{Transfer properties, Wick's theorem and vacuum averages for a linear quantum channel}%
\label{ch:T}
\subsection{Preliminary remarks}%
\label{ch:TP}
In the rest of the paper, we extend analyses of sections \ref{ch:O} and \ref{ch:H} to complex fields, bosonic and fermionic. This takes a few generalisations of the results of \mbox{Refs.\ \cite{API,APIII}}, which are the subject of this section. These generalisations may be divided into compulsory and elective. Compulsory generalisations are what in the introduction was called ``filling the gap:'' derivation of an analog of the test-case formula (\ref{eq:27VF}) for complex fields, absent in \cite{API,APIII}. The elective generalisation is an extension of complex fields to {\em linear quantum channels\/}. 

Quantum channel is a convenient formal way of making our analyses applicable to relativistic quantum fields with notrivial matrix structure. So, a conjugate pair of free fields in papers \cite{API,APIII} was understood as a Hermitian-conjugate pair, whereas the Dirac-adjoint of the four-component spin-half field (say) involves change of sign of two components. To cover such cases we make coefficient functions of the ``straight'' and ``adjoint'' fields independent, calling such pair a linear quantum channel. Conventional quantised fields, whether bosonic or fermionic, are special cases of quantum channels. 

The algebra of Green functions of a quantum channel is in essence that of a complex field (see appendix \ref{ch:A}). Results of \mbox{Refs.\ \cite{API,APIII}} may therefore be used for quantum channels with minimal amendments. It is ``filling the gap'' that takes most of the formal effort in this section. Extension to quantum channels reduces to a large extent to the observation that ``it does not change the algebra.'' 
\subsection{The channel operators}%
\label{ch:TD}
A linear quantum channel (linear channel, quantum channel, or just channel) is specified by a pair of ``tilde-conjugates,'' 
\settowidth{\auxl}{M}\setlength{\auxl}{0.1\auxl}
{\begin{align}{{
 \begin{aligned} 
{\hat\psi} (x,t) & = \sum_{\kappa}\sqrt{\frac{\hbar}{2\omega_{\kappa }}} 
\ensuremath{\big[
 u_{\kappa}(x) \hspace{1\auxl}\hat b_{\kappa} \hspace{1\auxl}\textrm{e}^{-i\omega _{\kappa}t}
+\tilde v_{\kappa}(x) \hspace{1\auxl}\hat c_{\kappa}^{\dag} \textrm{e}^{i\omega _{\kappa}t}
\big]} , \\
{\hat{\tilde\psi}}(x,t) & = \sum_{\kappa}\sqrt{\frac{\hbar}{2\omega_{\kappa }}} 
\ensuremath{\big[
 v_{\kappa}(x) \hspace{1\auxl} \hat c_{\kappa} \hspace{1\auxl}\textrm{e}^{-i\omega _{\kappa}t}
+\tilde u_{\kappa}(x) \hat b_{\kappa}^{\dag} \textrm{e}^{i\omega _{\kappa}t} 
\big]} , 
\end{aligned}}}%
\label{eq:57KC} 
\end{align}}%
where $ u_{\kappa}(x),\tilde v_{\kappa}(x), v_{\kappa}(x),\tilde u_{\kappa}(x)$ are four independent c-number functions. The meaning of the ``coordinate'' $x$ and of the ``mode index'' $\kappa $, as well as of the ``mode sum'' $\sum_{\kappa}$ and of the ``spatial integration'' $\int dx$, is problem-specific, cf.\ the remarks after eq.\ (\ref{eq:82UV}). The operators $\hat b_{\kappa},\hat b_{\kappa}^{\dag},\hat c_{\kappa},\hat c_{\kappa}^{\dag}$ form two standard creation/annihilation pairs for particles and antiparticles, which may be bosonic or fermionic, 
{\begin{align}{{
 \begin{aligned} 
\ensuremath{\big[
\hat b_{\kappa},\hat b_{\kappa'}^{\dag}
\big]}_{\mp} = 
\ensuremath{\big[
\hat c_{\kappa},\hat c_{\kappa'}^{\dag}
\big]}_{\mp} = \delta _{\kappa\kappa'}. 
\end{aligned}}}%
\label{eq:58KD} 
\end{align}}%
Here, $[\cdot,\cdot]_-$ stands for a commutator and $[\cdot,\cdot]_+$ for an anticommutator, 
{\begin{align}{{
 \begin{aligned} 
{[\hat b_{\kappa},\hat b_{\kappa'}^{\dag}]}_{\mp} 
= \hat b_{\kappa}\hat b_{\kappa'}^{\dag}\mp\hat b_{\kappa'}^{\dag}\hat b_{\kappa} ,
\end{aligned}}}%
\label{eq:67KP} 
\end{align}}%
etc. In eqs.\ (\ref{eq:58KD}), (\ref{eq:67KP}), the upper sign is for bosons and the lower for fermions. To avoid lengthy specifications, we shall also use blanket terms commutator, commute, etc., with respect to both commutational and anticommutational relations. So, we say that, except (\ref{eq:58KD}), the creation and annihilation\ operators pairwise commute, meaning that they commute for bosons and anticommute for fermions. 

The quantum channel is a natural generalisation of a complex quantised field, bosonic or fermionic. Equation (\ref{eq:57KC}) obviously alludes to Dirac's fermions \cite{Schweber}. In this case, the tilde stands for the Dirac adjoint, $x$ comprises the coordinate and the spinor index, $\kappa $ is a combination of the momentum and spinor index, and $ u_{\kappa}(x),\tilde v_{\kappa}(x), v_{\kappa}(x),\tilde u_{\kappa}(x)$ are suitable solutions of the Dirac equation. In general, these symbols may be assigned any meaning. 

Definitions (\ref{eq:57KC}) are {\em semirelativistic\/}: they assume existence of particles and antiparticles but not the relativistic covariance of the theory. The semirelativistic\ case\ includes relativistic theories but does not reduce to them. Nonrelativistic solid-state theories dealing with particles and holes are in our sense semirelativistic. 

\subsection{Response problem for the quantum channel}%
\label{ch:TF}
\subsubsection{The Hamiltonian}%
\label{ch:TFH}
The pair of ``modes'' (\ref{eq:57KC}) becomes a channel when one couples to them a pair of external sources. Formally, this is expressed by the Hamiltonian, 
{\begin{multline}\hspace{0.4\columnwidth}\hspace{-0.4\twocolumnwidth} 
{\begin{aligned}
& \hat H(t) = \hat H_0 + \hat H_{\text{I}}(t), \\ 
& \hat H_0 = \sum_{\kappa}\omega _{\kappa}\ensuremath{\big(
\hat b_{\kappa}^{\dag}\hat b_{\kappa} 
+ \hat c_{\kappa}^{\dag}\hat c_{\kappa}
 \big)} , \\ 
& \hat H_{\text{I}}(t) 
= -\int dx \hat H_{\text{I}}(x,t) 
\end{aligned}} 
\\ 
= -\int dx 
\ensuremath{\big[
{\hat{\tilde\psi}}(x,t)\sigma(x,t) + 
\tilde\sigma(x,t){\hat\psi}(x,t)
\big]} 
. 
\hspace{0.4\columnwidth}\hspace{-0.4\twocolumnwidth} 
\label{eq:75KX} 
\end{multline}}%
By definition, this Hamiltonian is written in the interaction picture, with $\hat H_0$ regarded free Hamiltonian, and $\hat H_{\text{I}}(t)$ interaction. The channel operators are interaction-picture operators. The corresponding Heisenberg operators are ${\hat\Psi}(x,t)$ and ${\hat{\tilde\Psi}}(x,t)$. 

\subsubsection{Summary of Grassmann fields}%
\label{ch:GV}
For bosons, the sources in (\ref{eq:75KX}) are c-numbers, for fermions, they are {\em Grassmann fields\/} \cite{APIII}. Unlike Beresin and Vasil'ev \cite{VasF,Beresin}, we define Grassmann fields as linear combinations of Grassmann generators, and the derivatives by them as anticommuting variational derivatives. To make the latter unique, we regard dimensionality of the underlying Grassmann algebra a variable parameter \cite{APIII}. For mathematical details see appendix \ref{ch:DGF}. 

For all practical purposes, Grassmann fields are formal quantities which commute with c-numbers and bosonic operators, 
and anticommute between themselves and with fermionic operators, e.g.,
{\begin{align}{{
 \begin{aligned} 
\sigma (x,t)\sigma (x',t') 
= \pm \sigma (x',t')\sigma (x,t), \\ 
\sigma (x,t)\tilde\sigma (x',t') 
= \pm \tilde\sigma (x',t')\sigma (x,t), \\ 
\sigma (x,t){\hat\psi} (x',t') 
= \pm {\hat\psi} (x',t')\sigma (x,t),
\end{aligned}}}%
\label{eq:13RV} 
\end{align}}%
etc. 
As in (\ref{eq:58KD}), the upper sign here is for bosons (when $\sigma $ and $\tilde\sigma $ are c-number fields), and the lower for fermions (when $\sigma $ and $\tilde\sigma $ are Grassmann fields). Differentiation by Grassmann variables follows simple rules, 
{\begin{align}{{
 \begin{aligned} 
& \frac{\delta \mathcal{C}}{\delta \sigma (x,t)} = 0, \\ 
& \frac{\delta \ensuremath{[
\mathcal{X} + \mathcal{Y}
]}}{\delta \sigma (x,t)}
= \frac{\delta 
\mathcal{X}
}{\delta \sigma (x,t)}
+ \frac{\delta 
\mathcal{Y}
}{\delta \sigma (x,t)}, \\ 
& {\begin{aligned}\frac{\delta \ensuremath{[
\sigma (x',t')\mathcal{X} 
]} }{\delta \sigma (x,t)} 
& = \delta (t-t')\delta (x,x')\mathcal{X} 
\\ & \ \ \ \pm \sigma (x',t')\frac{\delta \mathcal{X}}{\delta \sigma (x,t)} , \end{aligned}} 
\end{aligned}}}%
\label{eq:90QW} 
\end{align}}%
where $\mathcal{C}$ is an arbitrary quantity which does not depend on $\sigma (x,t)$, and $\mathcal{X}$, $\mathcal{Y}$ are arbitrary quantities which may or may not depend on $\sigma (x,t)$. The last of eqs.\ (\ref{eq:90QW}) holds for c-numbers with plus and for Grassmann fields with minus. Furthermore, for linear substitutions with c-number coefficients, chain laws are identical for c-number and Grassmann variational derivatives, cf.\ eq.\ (\ref{eq:96RC}) in appendix \ref{ch:TAD}. Together with the chain laws, eqs.\ (\ref{eq:13RV}) and (\ref{eq:90QW}) suffice for all algebraic manipulations in the paper. The rule of thumb is, watch the order of factors and how you commute/anticommute them, otherwise proceed as with c-numbers. 

\subsubsection{Transfer functions of the channel}%
\label{ch:TFF}
We are interested in the linear response functions of the quantum channel, 
{\begin{align}{{
 \begin{aligned} 
\Delta_{\text{R}}(x,x',t-t') & = \ensuremath{\bigg\langle 
\frac{\delta {\hat\Psi}(x,t)}{\delta \sigma (x',t')}
\bigg\rangle} \settoheight{\auxlv}{$\big|$}%
\raisebox{-0.3\auxlv}{$\big|_{\sigma =\tilde\sigma =0}$}, \\ 
\tilde\Delta_{\text{R}}(x',x,t'-t) & = \ensuremath{\bigg\langle 
\frac{\delta {\hat{\tilde\Psi}}(x,t)}{\delta \tilde\sigma (x',t')}
\bigg\rangle} \settoheight{\auxlv}{$\big|$}%
\raisebox{-0.3\auxlv}{$\big|_{\sigma =\tilde\sigma =0}$} . 
\end{aligned}}}%
\label{eq:14RW} 
\end{align}}%
They describe propagation of quantum signals in the channel from ${\hat{\tilde\psi}}$ to ${\hat\psi}$ and from ${\hat\psi}$ to ${\hat{\tilde\psi}}$, respectively. For a linear channel, these signals do not interfere with each other. That the derivative must be kept inside the average is a fermionic/Grassmann feature: Grassmann derivatives do not commute with quantum averaging (\mbox{Ref.\ \cite{APIII}}, section 4.2). 

Similar to the harmonic oscillator, we introduce only retarded Green functions, rather than more commonly used retarded and advanced functions \cite{Therm1}. 
It should not be overlooked that $\tilde\Delta_{\text{R}}$ is defined with a transposition of arguments. Such definition conveniently absorbs transpositions which otherwise occur in response transformations \cite{API,APII,APIII}. It also leads to natural algebraic structures in the case of fermions.
 
Quantum dynamics with Grassmann sources was considered, e.g., in \mbox{Ref.\ \cite{APIII}}, section 5. What matters is that, so far as we preserve the order of factors in $\hat H_{\text{I}}$, perturbation theory with this Hamiltonian may be constructed the usual way. It is then straightforward to obtain the explicit formulae, 
{\begin{align}{{
 \begin{aligned} 
\Delta_{\text{R}}(x,x',t-t') & 
= \frac{i}{\hbar }
\theta(t-t')\ensuremath{\big\langle 
\ensuremath{\big[
{\hat\psi}(x,t),{\hat{\tilde\psi}}(x',t')
\big]} _{\mp}
\big\rangle} , 
\end{aligned}}}%
\label{eq:76KY} 
\end{align}}%
and 
{\begin{align}{{
 \begin{aligned} 
\tilde\Delta_{\text{R}}(x',x,t'-t) 
= -\frac{i}{\hbar }
\theta(t-t')\ensuremath{\big\langle 
\ensuremath{\big[
{\hat\psi}(x',t'),{\hat{\tilde\psi}}(x,t)
\big]} _{\mp}
\big\rangle} , 
\end{aligned}}}%
\label{eq:78LA} 
\end{align}}%
cf. \mbox{Ref.\ \cite{APIII}}. 
Both in (\ref{eq:76KY}) and (\ref{eq:78LA}) the commutators are c-numbers, so that the quantum averaging is in fact redundant. The anomalous response functions of the channel vanish, 
{\begin{align}{{
 \begin{aligned} 
\ensuremath{\bigg\langle 
\frac{\delta {\hat\Psi}(x,t)}{\delta \tilde\sigma (x',t')}
\bigg\rangle} \settoheight{\auxlv}{$\big|$}%
\raisebox{-0.3\auxlv}{$\big|_{\sigma =\tilde\sigma =0}$}
= \ensuremath{\bigg\langle 
\frac{\delta {\hat{\tilde\Psi}}(x,t)}{\delta \sigma (x',t')}
\bigg\rangle} \settoheight{\auxlv}{$\big|$}%
\raisebox{-0.3\auxlv}{$\big|_{\sigma =\tilde\sigma =0}$} = 0 . 
\end{aligned}}}%
\label{eq:22SE} 
\end{align}}%
Details of the calculation, which may be interesting as an example of algebraic manipulation of Grassmann fields, are consigned to appendix \ref{ch:DER}.

Up to transposition of arguments and the $\pm i/\hbar $ factors, the transfer functions are the retarded and advanced parts of the same field commutator, and must therefore be related to each other. Indeed, in a real problem, the said commutator obeys some sort of free equation. Knowing its solution in one half of space-time suffices to find it in the other half. However, there does not seem to exist a closed relation between the two, valid for an arbitrary quantum channel. We treat $\Delta_{\text{R}}$ and $\tilde\Delta_{\text{R}}$ as if they were independent quantities. 

\begin{figure}[t]
\begin{center}
\includegraphics[width=230pt]{\Eps 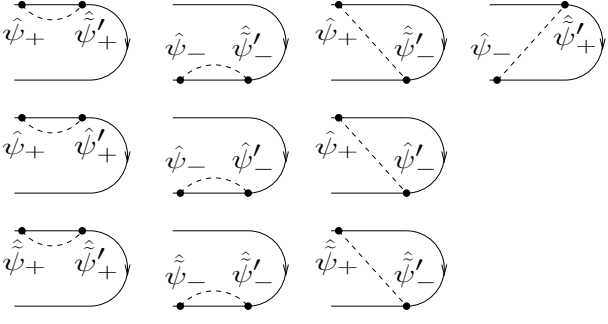}
\caption{The ten possible ways a pair of channel operators (bold dots) may be placed on the C-contour; $\hat \psi = \hat \psi (x,t)$, $\hat \psi' = \hat \psi (x',t')$, and similarly for the tilded operators. Dashed lines signify the corresponding contractions. Only those on the top row are nonzero.}
\label{fig:contr}
\end{center}
\end{figure}

\subsection{The Hori-Wick theorem for the quantum channel}%
\label{ch:W}
Definitions of the $T_C$ and normal orderings for bosonic and fermionic operators were discussed at length in \mbox{Ref.\ \cite{APIII}}, section 4.3; they are briefly reiterated in appendix \ref{ch:WO}. The Hori-Wick theorem for the quantum channel reads, 
{\begin{multline}\hspace{0.4\columnwidth}\hspace{-0.4\twocolumnwidth} 
 {\begin{aligned}
& {\begin{aligned}
& T_C\mathcal{F}(\hat \psi_+,{\hat{\tilde\psi}}_+,\hat \psi_-,{\hat{\tilde\psi}}_-) 
= {\mbox{\rm\boldmath$:$}}\mathcal{F}_N(\hat \psi ,{\hat{\tilde\psi}})
{\mbox{\rm\boldmath$:$}} , & 
\mathrm{where}\end{aligned}} 
\\ 
&\mathcal{F}_N( \psi_+,{\tilde\psi}_+ ) 
= \exp \mathcal{Z}_C\ensuremath{\bigg(
\frac{\delta }{\delta \psi_+},
\frac{\delta }{\delta {\tilde\psi}_+},
\frac{\delta }{\delta \psi_-},
\frac{\delta }{\delta {\tilde\psi}_-}
 \bigg)} 
\end{aligned}} 
\\ \times \mathcal{F}( \psi_+, {\tilde\psi}_+ ,\psi_-, {\tilde\psi}_-) 
\settoheight{\auxlv}{$|$}%
\raisebox{-0.3\auxlv}{$|_{\psi_-=\psi_+, {\tilde\psi}_-= {\tilde\psi}_+}$}. 
\hspace{0.4\columnwidth}\hspace{-0.4\twocolumnwidth} 
\label{eq:60PP} 
\end{multline}}%
In this formula, $\mathcal{F}(\cdot,\cdot,\cdot,\cdot)$ is an arbitrary c-number functional of four arguments. The reordering form in notation (\ref{eq:60TW}) is,
{\begin{multline}\hspace{0.4\columnwidth}\hspace{-0.4\twocolumnwidth} 
\mathcal{Z}_C\ensuremath{\big(
{\tilde f_+},
{f_+},
{\tilde f_-},
{f_-}
 \big)} 
= -i\varepsilon_{\mathrm{f}}\hbar \ensuremath{\big[
{\tilde f_+}\Delta _{\text{F}}{f_+}
-
{\tilde f_-}\tilde\Delta _{\text{F}}{f_-}
\\ 
+
{\tilde f_-}\Delta ^{(+)}{f_+}
-
{\tilde f_+}\Delta ^{(-)}{f_-}
\big]} 
, 
\hspace{0.4\columnwidth}\hspace{-0.4\twocolumnwidth} 
\label{eq:61PQ} 
\end{multline}}%
where 
{\begin{align}{{
 \begin{aligned} 
-i\hbar\Delta _{\text{F}}(x,x',t-t') 
& 
= \ensuremath{\big\langle 0\big|
T_+{\hat\psi} (x,t){\hat{\tilde\psi}} (x',t')
\big|0\big\rangle} 
, \\ 
i\hbar \tilde\Delta _{\text{F}}(x,x',t-t') 
&
= \ensuremath{\big\langle 0\big|
T_-{\hat\psi} (x,t){\hat{\tilde\psi}} (x',t')
\big|0\big\rangle} 
, \\ 
-i\hbar\Delta ^{(+)}(x,x',t-t') 
&
= \ensuremath{\big\langle 0\big|
{\hat\psi} (x,t){\hat{\tilde\psi}} (x',t')
\big|0\big\rangle} 
, \\ 
i\hbar\Delta ^{(-)}(x,x',t-t') 
&
= \varepsilon_{\mathrm{f}}\ensuremath{\big\langle 0\big|
{\hat{\tilde\psi}} (x',t'){\hat\psi} (x,t)
\big|0\big\rangle} , 
\end{aligned}}}%
\label{eq:90LP} 
\end{align}}%
are the four contractions characteristic of the channel. 
We introduced the fermionic sign factor, 
{\begin{align}{{
 \begin{aligned} 
\varepsilon_{\mathrm{f}} = 
\begin{cases}
+1 & {\mathrm{\ for\ bosons}}, \\ 
-1 & {\mathrm{\ for\ fermions}}.
\end{cases} 
\end{aligned}}}%
\label{eq:60FH} 
\end{align}}%
The $\pm$ symbol is hereafter reserved for purposes other than distinguishing bosons from fermions, in particular, for distinguishing the C-contour indices as in eq.\ (\ref{eq:9JN}) below; this notation applies throughout the rest of the paper. 
The $\psi $'s in (\ref{eq:60PP}) and the $f$'s in (\ref{eq:61PQ}) are auxiliary functional variables, c-numbers for bosons or Grassmann fields for fermions. 
Algebraic rules for Grassmann fields were summarised in section \ref{ch:GV}; for their justification see appendix \ref{ch:DGF}. Proof of eq.\ (\ref{eq:60PP}) is given in appendix \ref{ch:PW}.

Similar to the oscillator, {\em mere awareness\/} that proof of the Hori-Wick theorem (\ref{eq:60PP}) exists allows one to construct its ``mnemonic derivation'' in detailed analogy with section \ref{ch:OM}. An arbitrary quadratic form of four functional arguments is specified by ten kernels, which correspond to ten possible ways of placing a pair of channel operators on the C-contour, cf.\ fig.\ \ref{fig:contr}. Applying (\ref{eq:60PP}) to such pairs establishes the connection between kernels and contractions. This calculation utilizes a generalisation of eq.\ (\ref{eq:58WQ}) to complex fields, 
{\begin{align}{{
 \begin{aligned} 
\ensuremath{\bigg(
\frac{\delta }{\delta f}K\frac{\delta }{\delta \tilde f}
 \bigg)} f(x,t)\tilde f(x',t') = \varepsilon_{\mathrm{f}} K(x,x',t-t'), 
\end{aligned}}}%
\label{eq:59WR} 
\end{align}}%
where \mbox{$K(x,x',t-t')$} is a c-number kernel and \mbox{$f(x,t),\tilde f(x',t')$} are c-number functions (for bosons) or Grassmann fields (for fermions). Fermionic sign factors in (\ref{eq:61PQ}) and (\ref{eq:59WR}) cancel each other. Of ten contractions, only the four on the top row in fig.\ \ref{fig:contr} are nonzero. They are given by (\ref{eq:90LP}). 

\subsection{The test-case formula}%
\label{ch:CT}
As a test case when deriving response transformation of the Hori-Wick theorem we shall use a formula for the vacuum average, 
{\begin{multline}\hspace{0.4\columnwidth}\hspace{-0.4\twocolumnwidth} 
\Phi _{\textrm{vac}}\ensuremath{\big(
\tilde\eta_+,\eta_+,\tilde\eta_-,\eta_-
 \big)} 
\\ 
= \ensuremath{\big\langle 0\big|
T_C\exp\ensuremath{\big(
i\tilde\eta_+{\hat\psi}_+ 
+i{\hat{\tilde\psi}}_+\eta_+ 
-i\tilde\eta_-{\hat\psi}_-
-i{\hat{\tilde\psi}}_-\eta_-
 \big)}
\big|0\big\rangle} . 
\hspace{0.4\columnwidth}\hspace{-0.4\twocolumnwidth} 
\label{eq:52TN} 
\end{multline}}%
As for the oscillator, the {\em test-case formula\/} expresses calculation of $\Phi _{\textrm{vac}}$ by means of the Hori-Wick theorem, 
{\begin{multline}\hspace{0.4\columnwidth}\hspace{-0.4\twocolumnwidth} 
\Phi _{\textrm{vac}}\ensuremath{\big(
\tilde\eta_+,\eta_+,\tilde\eta_-,\eta_-
 \big)}
\\ 
= \exp \mathcal{Z}_C\ensuremath{\bigg(
\frac{\delta }{\delta \psi_+},
\frac{\delta }{\delta \tilde \psi_+},
\frac{\delta }{\delta \psi_-},
\frac{\delta }{\delta \tilde \psi_-}
 \bigg)}
\\ \times 
\exp\ensuremath{\big(
i\tilde\eta_+\psi_+ 
+i{\tilde\psi}_+\eta_+ 
-i\tilde\eta_-\psi_-
-i{\tilde\psi}_-\eta_-
 \big)}
\settoheight{\auxlv}{$|$}%
\raisebox{-0.3\auxlv}{$|_{\psi_{\pm}=\tilde \psi_{\pm}=0}$} 
\\ 
= \exp \ensuremath{\big[
\varepsilon_{\mathrm{f}}\mathcal{Z}_C (i\tilde\eta_+,i\eta_+,-i\tilde\eta_-,-i\eta_-) 
\big]} . 
\hspace{0.4\columnwidth}\hspace{-0.4\twocolumnwidth} 
\label{eq:55TR} 
\end{multline}}%
For bosons, the $\eta$'s and $\psi $'s are auxiliary c-number functions, for fermions, they are auxiliary Grassmann fields. The intermediate relation in (\ref{eq:55TR}) is found applying the Hori-Wick theorem to eq.\ (\ref{eq:52TN}) and using that, for any functional $\mathcal{F}_N$, 
{\begin{align}{{
 \begin{aligned} 
\ensuremath{\big\langle 0\big|
{\mbox{\rm\boldmath$:$}}\mathcal{F}_N\ensuremath{\big(
\hat \psi ,{\hat{\tilde\psi}} 
 \big)} 
{\mbox{\rm\boldmath$:$}}
\big|0\big\rangle} = \mathcal{F}_N\ensuremath{\big(
0,0
 \big)} . 
\end{aligned}}}%
\label{eq:54TQ} 
\end{align}}%
The final formula is now subject to verification. 

For bosons, proof of (\ref{eq:55TR}) is a straightforward generalisation of that of eq.\ (\ref{eq:57TT}) in section \ref{ch:OF}. The only difference is that we now have four shift operators to apply. To extend (\ref{eq:55TR}) to Grassmann fields, we note that the sign factor in the exponent there cancels the ``inbuilt'' sign factor in $\mathcal{Z}_C$, so that the last expression in eq.\ (\ref{eq:55TR}) actually reads, 
{\begin{multline}\hspace{0.4\columnwidth}\hspace{-0.4\twocolumnwidth} 
\Phi _{\textrm{vac}}\ensuremath{\big(
\tilde\eta_+,\eta_+,\tilde\eta_-,\eta_-
 \big)}
= \exp\ensuremath{\big\{
i\hbar \ensuremath{\big[
{\tilde \eta_+ }\Delta _{\text{F}} {\eta_+ }
-{\tilde \eta_- }\tilde\Delta _{\text{F}} {\eta_- }
\\ 
-{\tilde \eta_- }\Delta ^{(+)} {\eta_+ }
+{\tilde \eta_+ }\Delta ^{(-)} {\eta_- }
\big]}
\big\}} . 
\hspace{0.4\columnwidth}\hspace{-0.4\twocolumnwidth} 
\label{eq:74SZ} 
\end{multline}}%
This formula has been verified for c-numbers. To extend it to Grassmann fields, it suffices to prove that, in both cases, 
{\begin{multline}\hspace{0.4\columnwidth}\hspace{-0.4\twocolumnwidth} 
\Phi _{\textrm{vac}}\ensuremath{\big(
\tilde\eta_+,\eta_+,\tilde\eta_-,\eta_-
 \big)} \\ 
= F\ensuremath{\big(
{\tilde \eta_+ }\Delta _{\text{F}} {\eta_+ }, 
{\tilde \eta_- }\tilde\Delta _{\text{F}} {\eta_- }, 
{\tilde \eta_- }\Delta ^{(+)} {\eta_+ }, 
{\tilde \eta_+ }\Delta ^{(-)} {\eta_- }
 \big)} , 
\hspace{0.4\columnwidth}\hspace{-0.4\twocolumnwidth} 
\label{eq:75TA} 
\end{multline}}%
where $F(\cdot,\cdot,\cdot,\cdot)$ is a {\em unique\/} c-number function of four arguments. Equation (\ref{eq:75TA}) is proved in appendix \ref{ch:CF}. We therefore regard the test-case formula (\ref{eq:55TR}) proved for both bosons and fermions. 

\section{The causal Wick theorem in the semirelativistic\ case}%
\label{ch:S}
\subsection{Response transformation of contractions and the causal variables}%
\label{ch:SV}
As is shown in appendix \ref{ch:A}, the algebra of Green functions of a quantum channel is in essence that of a complex field, bosonic of fermionic \cite{API,APIII}. We may therefore use results of those papers with minimal generalisations. So, {\em response transformations\/} of contractions read, 
{\begin{align}{{
 \begin{aligned} 
\Delta^{(+)} (x,x',t-t') & = 
\Delta_{\text{R}}^{(+)}(x,x',t-t') \\ & \ \ \ -\tilde\Delta_{\text{R}}^{(+)}(x,x',t-t') , 
\\ 
\Delta^{(-)} (x,x',t-t') & = 
\Delta_{\text{R}}^{(-)}(x,x',t-t') \\ & \ \ \ -\tilde\Delta_{\text{R}}^{(-)}(x,x',t-t') , \\ 
\Delta _{\text{F}}(x,x',t-t') & =\Delta_{\text{R}}^{(+)}(x,x',t-t') 
\\ & \ \ \ + \tilde\Delta_{\text{R}}^{(-)}(x,x',t-t') , \\ 
\tilde\Delta _{\text{F}}(x,x',t-t') & =\Delta_{\text{R}}^{(-)}(x,x',t-t') 
\\ & \ \ \ + \tilde\Delta_{\text{R}}^{(+)}(x,x',t-t') . 
\end{aligned}}}%
\label{eq:60WS} 
\end{align}}%
These relations coincide with those derived in \mbox{Refs.\ \cite{API,APIII}}, except for two minor differences: there are now two (formally) independent response functions, $\Delta _{\text{R}}$ and $\tilde\Delta _{\text{R}}$, and $\tilde \Delta _{\text{R}}$ was defined with a transposition of arguments. 
 
Equations (\ref{eq:60WS}) hint at new functional variables, $\eta(x,t)$, $\tilde\eta(x,t)$, $\sigma_{\textrm{e}}(x,t)$ and $\tilde \sigma_{\textrm{e}}(x,t)$, such that, 
{\begin{align}{{
 \begin{aligned} 
\mathcal{Z}_C(i\tilde\eta_+,i\eta_+,-i\tilde\eta_-,-i\tilde\eta_-) 
= i \varepsilon_{\mathrm{f}} \ensuremath{\big(
\tilde\eta \Delta _{\text{R}}\sigma _{\textrm{e}} - \tilde\sigma _{\textrm{e}}\tilde \Delta _{\text{R}}\eta
 \big)} . 
\end{aligned}}}%
\label{eq:51UY} 
\end{align}}%
We continue using condensed notation (\ref{eq:60TW}). 
The bilinear forms in (\ref{eq:51UY}) clearly allude to the response properties of the quantum channel, see section \ref{ch:TF}. 
It should not be forgotten that evolution in $\tilde \Delta _{\text{R}}$ is from left to write, see eq.\ (\ref{eq:14RW}) and ensuing comments. 
The {\em causal variables\/} are introduced by the {\em response substitution\/} \cite{API,APII,APIII}, 
{\begin{align}{{
 \begin{aligned} 
\eta_{\pm} & = \frac{\sigma _{\textrm{e}}}{\hbar }\mp \eta ^{(\mp)}, & 
\tilde\eta_{\pm} & = \frac{\tilde\sigma _{\textrm{e}}}{\hbar }
\pm \tilde\eta ^{(\mp)}, 
\end{aligned}}}%
\label{eq:9JN} 
\end{align}}%
while the inverse substitution reads, 
{\begin{align}{{
 \begin{aligned} 
\eta & = 
\eta_--\eta_+
, & 
\sigma _{\textrm{e}} & = \hbar \ensuremath{\big[
\eta ^{(+)}_+ + \eta ^{(-)}_-
\big]} , \\ 
\tilde\eta & = 
\tilde\eta_+-\tilde\eta_-
, & 
\tilde\sigma _{\textrm{e}} & = \hbar \ensuremath{\big[
\tilde\eta ^{(+)}_+ + \tilde\eta ^{(-)}_-
\big]} , 
\end{aligned}}}%
\label{eq:5JJ} 
\end{align}}%
To verify (\ref{eq:51UY}), it suffices to apply eqs.\ (\ref{eq:90LP}) and use the relations, 
{\begin{align}{{
 \begin{aligned} 
\tilde f K^{(\pm)} f = \tilde f K f^{(\pm)} = \tilde f^{(\mp)} K f . 
\end{aligned}}}%
\label{eq:75UN} 
\end{align}}%
The latter follow from eqs.\ (\ref{eq:23KC}) in appendix \ref{ch:RTF}. 
For the reasons explained at the end of section \ref{ch:OR}, we moved the imaginary unit from the transformations (\ref{eq:9JN}), (\ref{eq:5JJ}) to the quadratic forms (\ref{eq:51UY}). Otherwise our equations follow phase conventions of Ref.\ \cite{APIII}, cf.\ also endnote \cite{endPh}. 
\subsection{Response transformation of the test-case formula}%
\label{ch:ST}
According to the schedule of section \ref{ch:HW}, we have completed the preliminary tasks 1--3, and are now approaching task 4. We wish to find functional variables $\zeta(x,t) ,\tilde\zeta(x,t) ,\psi _{\mathrm{e}}(x,t),\tilde\psi _{\mathrm{e}}(x,t)$ bringing the test-case formula to the form, 
{\begin{multline}\hspace{0.4\columnwidth}\hspace{-0.4\twocolumnwidth} 
\Phi _{\textrm{vac}}\ensuremath{\big(
\tilde\eta_+,\eta_+,\tilde\eta_-,\eta_-
 \big)}\settoheight{\auxlv}{$|$}%
\raisebox{-0.3\auxlv}{$|_{\textrm{c.v.}}$} 
\\ 
= \exp\ensuremath{\bigg(
- i\varepsilon_{\mathrm{f}} 
\frac{\delta }{\delta \psi_{\textrm{e}}} 
 \Delta _{\text{R}} 
\frac{\delta }{\delta \tilde\zeta }
\bigg)} 
\exp\ensuremath{\bigg( 
 i\varepsilon_{\mathrm{f}} 
\frac{\delta }{\delta \zeta } 
 \tilde\Delta _{\text{R}} 
\frac{\delta }{\delta\tilde \psi_{\textrm{e}}}
\bigg)} 
\\ \times 
\exp\ensuremath{\big(
i\tilde\eta \psi_{\textrm{e}} +i \tilde \zeta \sigma_{\textrm{e}} 
 \big)}
\exp\ensuremath{\big(
-i\tilde \psi_{\textrm{e}} \eta-i\tilde \sigma_{\textrm{e}}\zeta
 \big)}\settoheight{\auxlv}{$|$}%
\raisebox{-0.3\auxlv}{$|_{\psi_{\textrm{e}} = \tilde \psi_{\textrm{e}} = \zeta = \tilde \zeta =0 }$}
\\ 
= \exp\ensuremath{\big[
 i \tilde\eta \Delta _{\text{R}} \sigma_{\textrm{e}} 
- i \tilde \sigma_{\textrm{e}} \tilde\Delta _{\text{R}} \eta 
\big]} , 
\hspace{0.4\columnwidth}\hspace{-0.4\twocolumnwidth} 
\label{eq:94BC} 
\end{multline}}%
where c.v.\ refers to substitution (\ref{eq:9JN}). We split exponents so as to emphasise that 
eq.\ (\ref{eq:94BC}) in fact factorises. Factorisation is already seen in eqs.\ (\ref{eq:9JN}), which express two nearly identical transformations of the ``tilded'' and ``untilded'' variables. Transformation of the former coincides with (\ref{eq:33VN}), (\ref{eq:32VM}), while that of the latter differs in the sign of $\eta $. We therefore can recycle the algebra of section \ref{ch:OT} with minimal amendments. Postulating equality of the linear forms in (\ref{eq:55TR}) and in (\ref{eq:94BC}) we find the substitutions, 
{\begin{align}{{
 \begin{aligned} 
\psi _{\pm} & = {\psi _{\textrm{e}}}\mp {\hbar }\zeta ^{(\mp)}, & 
\tilde\psi _{\pm} & = {\tilde\psi _{\textrm{e}}}
\pm {\hbar }\tilde\zeta ^{(\mp)}, 
\end{aligned}}}%
\label{eq:69PY} 
\end{align}}%
where 
{\begin{align}{{
 \begin{aligned} 
\zeta 
& = \frac{\psi_--\psi_+}{\hbar }, 
& 
\psi_{\textrm{e}} 
& = \psi_+^{(+)} + \psi_-^{(-)} , 
\\
\tilde\zeta 
& = \frac{\tilde\psi_+-\tilde\psi_-}{\hbar }, 
& 
\tilde\psi_{\textrm{e}} 
& = \tilde\psi_+^{(+)} + \tilde\psi_-^{(-)} . 
\end{aligned}}}%
\label{eq:68PX} 
\end{align}}%
Similar to the real field, these transformations are rescaled versions of (\ref{eq:9JN}), (\ref{eq:5JJ}). Comparing them to eqs.\ (\ref{eq:38VT}) we see that transformation of the ``tilded'' variables is identical to (\ref{eq:38VT}), while that of the ``untilded'' ones differs in the sign of $\zeta $. Transformations laws for the derivatives by ``tilded'' variables are therefore identical to (\ref{eq:34KQ}), while for the ``untilded'' ones we have to adapt (\ref{eq:34KQ}) to the said sign change, 
{\begin{align}{{
 \begin{aligned} 
\mp i \frac{\delta }{\delta \psi_{\pm}} 
& = \frac{i}{\hbar } \frac{\delta }{\delta \zeta} 
\mp i \ensuremath{\bigg(
\frac{\delta }{\delta \psi_{\textrm{e}}}
 \bigg)} ^{(\mp)} , 
\\ 
\mp i \frac{\delta }{\delta \tilde \psi_{\pm}} 
& = -\frac{i}{\hbar } \frac{\delta }{\delta \tilde\zeta} 
\mp i \ensuremath{\bigg(
\frac{\delta }{\delta \tilde \psi_{\textrm{e}}}
 \bigg)} ^{(\mp)} . 
\end{aligned}}}%
\label{eq:76UP} 
\end{align}}%
In continuing analogy with the real field, these transformations emulate substitution (\ref{eq:9JN}) with the replacements, 
{\begin{align}{{
 \begin{aligned} 
\eta_{\pm} & \to \mp i \frac{\delta }{\delta \tilde \psi_{\pm}}, & 
\tilde \eta_{\pm} & \to \mp i \frac{\delta }{\delta \psi_{\pm}}, \\ 
\sigma_{\textrm{e}} & \to -i \frac{\delta }{\delta \tilde\zeta}, & 
\tilde \sigma_{\textrm{e}} & \to i \frac{\delta }{\delta \zeta}, \\ 
\eta & \to i\frac{\delta }{\delta \tilde \psi_{\textrm{e}}}, & 
\tilde \eta & \to -i\frac{\delta }{\delta \psi_{\textrm{e}}} . 
\end{aligned}}}%
\label{eq:75VZ} 
\end{align}}%
so that the transformation of the quadratic forms of derivatives reduces in fact to eq.\ (\ref{eq:51UY}). This concludes step 4 of the schedule of section \ref{ch:HW} for quantum channels. 
\subsection{The causal Wick theorem}%
\label{ch:SW}
Step 5 of the aforementioned schedule is straightforward. Applying transformations (\ref{eq:69PY}) to the formula for $\mathcal{F}_N$ in the Hori-Wick theorem (\ref{eq:60PP}) we find, 
\begin{widetext} 
{\begin{multline}
\mathcal{F}_N\ensuremath{\big(
\psi _{\mathrm{e}},\tilde\psi _{\mathrm{e}}
 \big)} =
\exp\ensuremath{\bigg(
- i\varepsilon_{\mathrm{f}} 
\frac{\delta }{\delta \psi_{\textrm{e}}} 
 \Delta _{\text{R}} 
\frac{\delta }{\delta \tilde\zeta }
+ i\varepsilon_{\mathrm{f}} 
\frac{\delta }{\delta \zeta } 
 \tilde\Delta _{\text{R}} 
\frac{\delta }{\delta\tilde \psi_{\textrm{e}}}
\bigg)} 
\\ \times 
\mathcal{F}\ensuremath{\big(
\psi_{\mathrm{e}}-\hbar \zeta ^{(-)} , 
\tilde\psi_{\mathrm{e}}+\hbar \tilde\zeta ^{(-)} , 
\psi_{\mathrm{e}}+\hbar \zeta ^{(+)} , 
\tilde\psi_{\mathrm{e}}-\hbar \tilde\zeta ^{(+)} 
 \big)} 
\settoheight{\auxlv}{$\big|$}%
\raisebox{-0.3\auxlv}{$\big|_{\zeta =\tilde\zeta =0}$} \, .
\label{eq:61WT} 
\end{multline}}%
\end{widetext}%
The first of eqs.\ (\ref{eq:60PP}) remains unchanged. Equation (\ref{eq:61WT}) generalises eq.\ (\ref{eq:94BC}) to operators and {\em ipso facto\/} to arbitrary quantum states of the field (channel). 

\section{The causal Wick theorem in the nonrelativistic\ case}%
\label{ch:N}
\subsection{Motivation}%
\label{ch:NR}
The nonrelativistic\ case\ of quantum channel follows by dropping antiparticles (holes) in the channel operators (\ref{eq:57KC}), 
that is, formally, by setting, 
{\begin{align}{{
 \begin{aligned} 
\tilde v_{\kappa}(x)= v_{\kappa}(x)=0
. 
\end{aligned}}}%
\label{eq:71XD} 
\end{align}}%
This also covers a physically important case of quantum electrodynamics under the resonance, or {\em rotating wave\/}, approximation (RWA), commonly used in spectroscopy and quantum optics. Under the RWA, one works not with the real field operator (\ref{eq:82UV}), but with its frequency-positive and negative\ parts, 
{\begin{align}{{
 \begin{aligned} 
\hat Q^{(+)} (x,t)
& = \sum_{\kappa}\sqrt{\frac{\hbar}{2\omega_{\kappa }}} 
u_{\kappa}(x) \hat a_{\kappa} \textrm{e}^{-i\omega _{\kappa}t } , \\
\hat Q^{(-)} (x,t)
& = \sum_{\kappa}\sqrt{\frac{\hbar}{2\omega_{\kappa }}} 
u_{\kappa}^*(x) \hat a_{\kappa}^{\dag} \textrm{e}^{i\omega _{\kappa}t }
, 
\end{aligned}}}%
\label{eq:62WU} 
\end{align}}%
Up to notation, 
{\begin{align}{{
 \begin{aligned} 
& \hat\psi(x,t) = \hat Q^{(+)} (x,t), & 
& {\hat{\tilde\psi}}(x,t) = \hat Q^{(-)} (x,t), 
\end{aligned}}}%
\label{eq:63WV} 
\end{align}}%
eqs.\ (\ref{eq:62WU}) define a nonrelativistic\ quantum channel. Interactions (\ref{eq:65WX}) generalised to the real field and (\ref{eq:75KX}) for the channel coincide under the RWA, 
{\begin{multline}\hspace{0.4\columnwidth}\hspace{-0.4\twocolumnwidth} 
\int dx \hat Q(x,t)J_{\mathrm{e}}(x,t) \\ 
= 
\int dx \ensuremath{\big[
\tilde\sigma(x,t) {\hat\psi}(x,t) + 
\sigma(x,t) {\hat{\tilde\psi}}(x,t)
\big]} , 
\hspace{0.4\columnwidth}\hspace{-0.4\twocolumnwidth} 
\label{eq:67WZ} 
\end{multline}}%
subject to another instance of notation, 
{\begin{align}{{
 \begin{aligned} 
& \sigma (x,t) = J_{\mathrm{e}}^{(+)} (x,t), & 
& \tilde\sigma (x,t) = J_{\mathrm{e}}^{(-)} (x,t) . 
\end{aligned}}}%
\label{eq:68XA} 
\end{align}}%
This leads to redefinition of the response function by the first of eqs.\ (\ref{eq:14RW}), and of the contractions by eqs.\ (\ref{eq:90LP}). 

Heed should be taken of the fact that redefinition of the response function under the RWA implies that we do not regard $\sigma (x,t)$ and $\tilde\sigma (x,t)$ as being, respectively, frequency-positive\ and frequency-negative. Indeed, Kubo's theory assumes that no conditions are imposed on the sources. However, with \mbox{$
\sigma (x,t)
$} and \mbox{$
\tilde\sigma (x,t)
$} constrained to the subspaces of frequency-positive and negative\ functions, we must also use differentiation with constrains. The latter is defined for an arbitrary functional $\mathcal{F}(\cdot)$ as \cite{ConstrD}, 
{\begin{multline}\hspace{0.4\columnwidth}\hspace{-0.4\twocolumnwidth} 
\frac{\delta \mathcal{F}\ensuremath{\big(
J_{\mathrm{e}}
 \big)} }{\delta J_{\mathrm{e}}^{(\pm)}(x,t)} 
\\ 
= 
\int dt' \delta ^{(\pm)}(t'-t)\frac{\delta \mathcal{F}\ensuremath{\big(
J_{\mathrm{e}}
 \big)}}{\delta J_{\mathrm{e}}(x,t')} 
= 
\ensuremath{\bigg[
\frac{\delta \mathcal{F}\ensuremath{\big(
J_{\mathrm{e}}
 \big)}}{\delta J_{\mathrm{e}}(x,t)}
\bigg]} ^{(\mp)}. 
\hspace{0.4\columnwidth}\hspace{-0.4\twocolumnwidth} 
\label{eq:70XC} 
\end{multline}}%
Irrespective of any particulars, such constrained differentiation cannot lead to causal (retarded) response because of the final separation of the frequency-positive and negative\ parts. We return to the discussion elsewhere. 

A somewhat confusing feature of the nonrelativistic\ case\ is that, {\em physically\/}, it emerges as an {\em approximation\/} to more complicated cases, while, {\em formally\/}, it comprises a coherent set of {\em exact\/} relations. As was just illustrated, this exactness is a consequence of a consistent set of approximations (see also a remark at the end of appendix B.4 in \cite{APIII}). A physically meaningful discussion of this point is postponed till particular models are approached. In this paper we treat the nonrelativistic\ case\ as a mere formal consequence of eq.\ (\ref{eq:71XD}). 

\subsection{Response transformations of contractions for a nonrelativistic\ quantum channel}%
\label{ch:TNR}
Owing to (\ref{eq:71XD}), the frequency-negative\ part of the commutator vanishes, 
{\begin{align}{{
 \begin{aligned} 
\Delta ^{(-)}(x,x',t-t') = 0 . 
\end{aligned}}}%
\label{eq:50TL} 
\end{align}}%
The algebra of Green functions, expressed by eqs.\ (\ref{eq:94LT})--(\ref{eq:92LR}), (\ref{eq:93LS}) in appendix \ref{ch:A} then becomes, 
{\begin{align}{{
 \begin{aligned} 
\Delta_{\text{F}}(x,x',t-t') & =\Delta_{\text{R}}(x,x',t-t') \\ & \ \ \ 
= \theta(t-t')\Delta^{(+)}(x,x',t-t'), \\ 
\tilde\Delta_{\text{F}}(x,x',t-t') & = \tilde\Delta_{\text{R}}(x,x',t-t') \\ & \ \ \ 
= -\theta(t'-t)\Delta^{(+)}(x,x',t-t'), \\ 
\Delta^{(+)} (x,x',t-t') & =\Delta_{\text{R}}(x,x',t-t') -\tilde\Delta_{\text{R}}(x,x',t-t') . 
\end{aligned}}}%
\label{eq:3MB} 
\end{align}}%
Equations (\ref{eq:3MB}) already contain the expression for $\Delta_{\text{F}}$, $\tilde\Delta_{\text{F}}$ and $\Delta^{(+)}$ by $\Delta_{\text{R}}$ and $\tilde\Delta_{\text{R}}$. No additional transformation is necessary. 

\subsection{The test-case formula\ldots}%
\label{ch:NV}
The nonrelativistic\ case\ is not covered in our \mbox{Refs.\ \cite{API,APIII}} (it was considered in \mbox{Ref.\ \cite{BWO}}, but none of the formulae we need were derived there). The good news is that, formal differences notwithstanding, the key relations coincide in the semirelativistic\ and nonrelativistic\ case s. We therefore recycle as much as possible of the semirelativistic\ case, mostly marking the differences. So, the Hori-Wick theorem (\ref{eq:60PP}) and the test-case formulae (\ref{eq:52TN}), (\ref{eq:55TR}) equally apply in the nonrelativistic\ case. We can also recycle eq.\ (\ref{eq:51UY}) postulating causal variables. 
The definition of the causal variables is already different. 

Satisfying eq.\ (\ref{eq:51UY}) in the nonrelativistic\ case\ is straightforward. Insertion of eqs.\ (\ref{eq:3MB})) into the LHS of (\ref{eq:51UY}) yields, 
{\begin{multline}\hspace{0.4\columnwidth}\hspace{-0.4\twocolumnwidth} 
\mathcal{Z}_C(i\tilde\eta_+,i\eta_+,-i\tilde\eta_-,-i\tilde\eta_-) \\ 
= i\varepsilon_{\mathrm{f}}\hbar \ensuremath{\big[
\ensuremath{\big(
\tilde \eta_+ - 
\tilde \eta_-
 \big)}
\Delta_{\text{R}}\eta_+ 
-
\tilde \eta_-
\tilde\Delta_{\text{R}}
\ensuremath{\big(
\eta_- - \eta_+
 \big)} 
\big]} . 
\hspace{0.4\columnwidth}\hspace{-0.4\twocolumnwidth} 
\label{eq:53VA} 
\end{multline}}%
Comparing this to (\ref{eq:51UY}) we see that the latter is safisfied by the variables, (omitting arguments for brevity)
{\begin{align}{{
 \begin{aligned} 
\eta & = \eta_- -\eta_+, & 
\sigma_{\textrm{e}} & = \hbar \eta_+, \\ 
\tilde\eta & = \tilde \eta_+ -\tilde \eta_-, & 
\tilde \sigma_{\textrm{e}} & = \hbar \tilde \eta_- . 
\end{aligned}}}%
\label{eq:54VB} 
\end{align}}%
Inverting (\ref{eq:54VB}) we find the {\em response substitution\/}, 
{\begin{align}{{
 \begin{aligned} 
\eta_+ & = \frac{\sigma_{\textrm{e}}}{\hbar }, & 
\tilde\eta_+ & = \tilde\eta + \frac{\tilde \sigma_{\textrm{e}}}{\hbar }, \\ 
\tilde\eta_- & = \frac{\tilde \sigma_{\textrm{e}}}{\hbar }, & 
\eta_- & = \eta + \frac{\sigma_{\textrm{e}}}{\hbar }. 
\end{aligned}}}%
\label{eq:56VD} 
\end{align}}%
Unlike eqs.\ (\ref{eq:9JN}), (\ref{eq:5JJ}), eqs.\ (\ref{eq:54VB}), (\ref{eq:56VD}) are algebraic and not integral transformations. 

\subsection{\ldots and its response transformation}%
\label{ch:NT}
The next relation we reuse literally is eq.\ (\ref{eq:94BC}). Identity of the linear forms in (\ref{eq:55TR}) and in (\ref{eq:94BC}) requires that, 
{\begin{multline}\hspace{0.4\columnwidth}\hspace{-0.4\twocolumnwidth} 
i\tilde\eta_+ \psi _+ + 
i\tilde\psi _+ \eta _+ - 
i\tilde\eta_- \psi _- - 
i\tilde\psi _- \eta _- 
\\ 
= 
i\tilde\eta \psi _+ 
+ i\frac{\tilde\psi _+-\tilde\psi _-}{\hbar }\sigma _{\textrm{e}} 
- i\tilde\psi _-\eta 
- i\tilde\sigma _{\textrm{e}}\frac{\psi _--\psi _+}{\hbar } 
, 
\hspace{0.4\columnwidth}\hspace{-0.4\twocolumnwidth} 
\label{eq:73UL} 
\end{multline}}%
where we applied substitution (\ref{eq:56VD}) and rearranged the terms. This formula hints at the variables, 
{\begin{align}{{
 \begin{aligned} 
\zeta & = \frac{\psi _--\psi _+}{\hbar }, & 
\psi _{\textrm{e}} & = \psi_+, \\ 
\tilde\zeta & = \frac{\tilde\psi _+-\tilde\psi _-}{\hbar }, & 
\tilde \psi _{\textrm{e}} & = \tilde \psi_- , 
\end{aligned}}}%
\label{eq:69UF} 
\end{align}}%
with the inverse substitution being, 
{\begin{align}{{
 \begin{aligned} 
\psi _+ & = \psi _{\mathrm{e}}, & 
\tilde\psi _+ & = \tilde\psi _{\mathrm{e}} +\hbar \tilde\zeta . 
\\ 
\tilde\psi _- & = \tilde\psi _{\mathrm{e}}, 
& 
\psi _- & = \psi _{\mathrm{e}} +\hbar \zeta . 
\end{aligned}}}%
\label{eq:72XE} 
\end{align}}%
Equations (\ref{eq:54VB}), (\ref{eq:69UF}) and (\ref{eq:56VD}), (\ref{eq:72XE}) are rescaled versions of each other. By making use of the standard chain rules we find, 
{\begin{align}{{
 \begin{aligned} 
-i\frac{\delta }{\delta \tilde \psi_+} 
& = -\frac{i}{\hbar }\,\frac{\delta }{\delta \tilde\zeta}, & 
-i\frac{\delta }{\delta \psi_+} 
& = -i\frac{\delta }{\delta \psi_{\textrm{e}}} 
+ \frac{i}{\hbar }\,\frac{\delta }{\delta \zeta}, \\
i\frac{\delta }{\delta \psi_-} 
& = \frac{i}{\hbar }\,\frac{\delta }{\delta \zeta}, & 
i\frac{\delta }{\delta \tilde \psi_-} 
& = i\frac{\delta }{\delta \tilde \psi_{\textrm{e}}} 
- \frac{i}{\hbar }\,\frac{\delta }{\delta \tilde\zeta} . 
\end{aligned}}}%
\label{eq:74VY} 
\end{align}}%
We remind that for linear transformations with c-number coefficients the chain rules are identical for c-number and Grassmann variational derivatives, see appendix \ref{ch:TAD}. As in the semirelativistic\ case, the purpose of additional phase factors in (\ref{eq:74VY}) is to emulate response substitution (\ref{eq:56VD}). Indeed, the latter is found from the former by replacements (\ref{eq:75VZ}); that is yet another formula we reuse literally. 
Such replacements turn eq.\ (\ref{eq:51UY}) with $\Delta ^{(-)}=0$ into, 
{\begin{multline}\hspace{0.4\columnwidth}\hspace{-0.4\twocolumnwidth} 
\mathcal{Z}_C\ensuremath{\bigg(
\frac{\delta }{\delta \psi_+},
\frac{\delta }{\delta \tilde \psi_+},
\frac{\delta }{\delta \psi_-},
\frac{\delta }{\delta \tilde \psi_-}
 \bigg)} \\ = -i\varepsilon_{\mathrm{f}}\ensuremath{\bigg(
\frac{\delta }{\delta \psi_{\textrm{e}}}
\Delta_{\text{R}} 
\frac{\delta }{\delta \tilde\zeta }
- 
\frac{\delta }{\delta \zeta } 
\tilde\Delta_{\text{R}} 
\frac{\delta }{\delta\tilde \psi_{\textrm{e}}}
 \bigg)} , 
\hspace{0.4\columnwidth}\hspace{-0.4\twocolumnwidth} 
\label{eq:76WA} 
\end{multline}}%
proving that variables (\ref{eq:69UF}) are indeed the right set of causal variables. 
\subsection{The causal Wick theorem}%
\label{ch:NW}
Applying transformartions (\ref{eq:72XE}) to the Hori-Wick theorem (\ref{eq:60PP}) we find the nonrelativistic\ counterpart of eq.\ (\ref{eq:61WT}), 
{\begin{align}{{
 \begin{aligned} 
\mathcal{F}_N\ensuremath{\big(
\psi _{\mathrm{e}},\tilde\psi _{\mathrm{e}}
 \big)} =
\exp\ensuremath{\bigg(
- i\varepsilon_{\mathrm{f}} 
\frac{\delta }{\delta \psi_{\textrm{e}}} 
 \Delta _{\text{R}} 
\frac{\delta }{\delta \tilde\zeta }
+ i\varepsilon_{\mathrm{f}} 
\frac{\delta }{\delta \zeta } 
 \tilde\Delta _{\text{R}} 
\frac{\delta }{\delta\tilde \psi_{\textrm{e}}}
\bigg)} 
\\ \times 
\mathcal{F}\ensuremath{\big(
\psi_{\mathrm{e}} , 
\tilde\psi_{\mathrm{e}}+\hbar \tilde\zeta , 
\psi_{\mathrm{e}}+\hbar \zeta , 
\tilde\psi_{\mathrm{e}} 
 \big)} 
\settoheight{\auxlv}{$\big|$}%
\raisebox{-0.3\auxlv}{$\big|_{\zeta =\tilde\zeta =0}$} \, .
\end{aligned}}}%
\label{eq:73XF} 
\end{align}}%
The first of eqs.\ (\ref{eq:60PP}) remains intact. It should not be forgotten that eq.\ (\ref{eq:73XF}) is subject to condition (\ref{eq:50TL}), which in turn follows from (\ref{eq:71XD}). We have thus covered all conceivable cases of interest: the real, the semirelativistic, and the nonrelativistic\ ones. 

\section{Conclusion and outlook}%
\label{ch:CO4}
We have identified transformations of the functional (Hori's) form of Wick's theorem for operators which bring out its underlying physical content: propagation of free fields in space and time. The resulting relations are called {\em causal Wick theorems\/}. They are the main result of this paper. Their full power shows up in problems involving nonlinear interactions. In particular, they will be instrumental in understanding electromagnetic actions and back-actions of distinguishable macroscopic devices, which will be the subject of forthcoming papers. 
\section{Acknowledgements}%
Support of SFB/TRR 21 and of the Humboldt Foundation is gratefully acknowledged. 
\appendix
\section{Formal definitions}%
\label{ch:DEF}
\subsection{Operator orderings}%
\label{ch:WO}
For fermionic operators, both time and normal orderings are defined with the sign factor (\ref{eq:60FH}). So, for the normal ordering, 
{\begin{align}{{
 \begin{aligned} 
{\mbox{\rm\boldmath$:$}}
\hat b_{\kappa}^{\dag}\hat b_{\kappa'}
{\mbox{\rm\boldmath$:$}} & = \varepsilon_{\mathrm{f}}{\mbox{\rm\boldmath$:$}}
\hat b_{\kappa'}\hat b_{\kappa}^{\dag}
{\mbox{\rm\boldmath$:$}} = \hat b_{\kappa'}^{\dag} \hat b_{\kappa}, \\ 
{\mbox{\rm\boldmath$:$}}
\hat c_{\kappa}^{\dag}\hat b_{\kappa'}
{\mbox{\rm\boldmath$:$}} & = \varepsilon_{\mathrm{f}}{\mbox{\rm\boldmath$:$}}
\hat b_{\kappa'}\hat c_{\kappa}^{\dag}
{\mbox{\rm\boldmath$:$}} = \hat b_{\kappa'}\hat c_{\kappa}^{\dag} , 
\end{aligned}}}%
\label{eq:60KF} 
\end{align}}%
and so on, cf.\ eqs.\ (\ref{eq:57KC}). Similarly, for the forward and reverse time orderings, 
{\begin{align}{{
 \begin{aligned} 
T_+{\hat\psi} (x,t){\hat{\tilde\psi}}(x',t') & = 
\theta(t-t'){\hat\psi} (x,t){\hat{\tilde\psi}}(x',t')
\\ & \ \ \ +\varepsilon_{\mathrm{f}}\theta(t'-t){\hat{\tilde\psi}}(x',t'){\hat\psi} (x,t), \\ 
T_-{\hat\psi} (x,t){\hat{\tilde\psi}}(x',t') & = 
\theta(t'-t){\hat\psi} (x,t){\hat{\tilde\psi}}(x',t')
\\ & \ \ \ +\varepsilon_{\mathrm{f}}\theta(t-t'){\hat{\tilde\psi}}(x',t'){\hat\psi} (x,t), \\ 
\end{aligned}}}%
\label{eq:61KH} 
\end{align}}%
etc. Sign rules for ordered fermionic products were discussed at length in \mbox{Ref.\ \cite{APIII}}. They reduce, in fact, to two statements. Firstly, under the orderings bosonic operators behave as c-numbers (i.e., commute with any other operator), while fermionic operators behave as Grassmann fields (i.e., anticommute with fermionic operators and Grassmann fields, and commute with bosonic operators). Secondly, the order of factors ``as written'' acquires no sign factor (the rule of visual consistency). These rules are evident in eqs.\ (\ref{eq:60KF}) and (\ref{eq:61KH}). 

In formulae, the $T_{\pm}$-orderings mostly occur as {\em double time ordered\/} structures $T_-\cdots T_+\cdots$. Rather than visually keeping the operators under the $T_{\pm}$-orderings, one marks the operators with the ${}_{\pm}$ indices and allows them to commute freely. For instance, 
{\begin{multline}\hspace{0.4\columnwidth}\hspace{-0.4\twocolumnwidth} 
T_-{\hat\psi}(t){\hat{\tilde\psi}}(t')\, 
T_+{\hat\psi}(t''){\hat{\tilde\psi}}(t''') 
\\ 
= T_C{\hat\psi}_-(t){\hat{\tilde\psi}}_-(t') 
{\hat\psi}_+(t''){\hat{\tilde\psi}}_+(t''') 
, 
\hspace{0.4\columnwidth}\hspace{-0.4\twocolumnwidth} 
\label{eq:23SF} 
\end{multline}}%
and so on.
The ${}_{\pm}$ indices serve only for ordering purposes and otherwise should be disregarded. 
Using $T_C$ results in a reduction of formulae in bulk which may be truly dramatic (compare, e.g., expressions for response functions in \cite{APII} and in \cite{APIII}). The rule of visual consistency is implicit in (\ref{eq:23SF}), where the order of factors is identical on both sides. Sign rules for more complicated cases are evident from the example, 
{\begin{multline}\hspace{0.4\columnwidth}\hspace{-0.4\twocolumnwidth} 
T_C{\hat\psi}_+(t){\hat{\tilde\psi}}_-(t'){\hat\psi}_+(t'') = 
\varepsilon_{\mathrm{f}} T_C{\hat{\tilde\psi}}_-(t'){\hat\psi}_+(t){\hat\psi}_+(t'') \\ 
= \varepsilon_{\mathrm{f}} {\hat{\tilde\psi}}(t')
T_+{\hat\psi}(t){\hat\psi}(t'') = 
{\hat{\tilde\psi}}(t')T_+{\hat\psi}(t''){\hat\psi}(t) . 
\hspace{0.4\columnwidth}\hspace{-0.4\twocolumnwidth} 
\label{eq:24SH} 
\end{multline}}%
At the last step here, the operators under the $T_+$ ordering were formally commuted so as to absorb the sign factor.

The $T_C$-ordering is commonly visualised as an ordering on the so-called C-contour depicted in Fig.\ \ref{fig:C}. It travels firstly from $t=-\infty$ to $t=+\infty$ (forward branch) and then back to $t=-\infty$ (reverse branch). One visualises operators marked with pluses and minuses as positioned, respectively, on the forward and reverse branches. The ordering rule for a $T_C$-product is consistent with the travelling order of the C-contour. 

We note in passing that, in general, equations like (\ref{eq:61KH}) are mathematically troublesome. In nonlinear problems, Heisenberg operators are, strictly speaking, operator-valued singular functions. Multiplying them by step-functions requires specifications. Even in the simplest one-mode case \cite{BWO}, a {\em causal regularisation\/} was necessary to fix the stochastic calculus (Ito) in phase-space equations. Divergences and consequent regularisation/renormalisation procedures in quantum field theory \cite{Schweber} are of the same origin.
\subsection{The frequency-positive and negative\ parts of functions}%
\label{ch:RTF}
Separating the frequency-positive and negative\ parts of a function is defined as, 
{\begin{align} 
& f(t) = f^{(+)}(t) + f^{(-)}(t) , 
\label{eq:82RZ} 
\\ 
& f^{(\pm)}(t) = \int \frac{d\omega }{2\pi }\textrm{e}^{-i\omega t}\theta(\pm\omega )
f_{\omega }, & 
& f_{\omega } = \int dt \textrm{e}^{i\omega t}f(t) . 
\label{eq:4JH} 
\end{align}}%
The frequency-positive and negative\ parts are always defined with respect to native arguments of functions, e.g.,
{\begin{align}{{
 \begin{aligned} 
f^{(\pm)}(t'-t) = f^{(\pm)}(t)\settoheight{\auxlv}{$|$}%
\raisebox{-0.3\auxlv}{$|_{t\to t'-t}$} \, . 
\end{aligned}}}%
\label{eq:49TK} 
\end{align}}%
The ${}^{(\pm)}$ operations are conveniently expressed as integral transformations, 
{\begin{align}{{
 \begin{aligned} 
f^{(\pm)}(t) = \int dt' \delta ^{(\pm)}(t-t') f(t'), 
\end{aligned}}}%
\label{eq:39KV} 
\end{align}}%
where 
{\begin{align}{{
 \begin{aligned} 
\delta ^{(\pm)}(t) = \delta ^{(\mp)}(-t) = \ensuremath{\pm\frac{1}{2\pi i(t\mp i0^+)}} 
\end{aligned}}}%
\label{eq:40KW} 
\end{align}}%
are the frequency-positive and negative\ parts of the delta-function. Of use are the formulae, 
{\begin{gather}{{
 \begin{gathered} 
\int dt f^{(\pm)}(t) g(t) = \int dt f(t) g^{(\mp)}(t) 
= \int dt f^{(\pm)}(t) g^{(\mp)}(t) , \\ 
\int dt f^{(+)}(t) g^{(+)}(t) = \int dt f^{(-)}(t) g^{(-)}(t) = 0 , 
\end{gathered}}}%
\label{eq:23KC} 
\end{gather}}%
where $f(t)$ and $g(t)$ are arbitrary functions. They may be verified using eq.\ (\ref{eq:4JH}) and the standard equality for Fourier-transforms, 
{\begin{align}{{
 \begin{aligned} 
\int dt f(t) g(t) = \int \frac{d\omega }{2\pi } f_{\omega } g_{-\omega } . 
\end{aligned}}}%
\label{eq:25KE} 
\end{align}}%
For more details see \mbox{Ref.\ \cite{APII}}, appendix A. 
\section{Variational derivatives by Grassmann fields}%
\label{ch:DGF}
\subsection{Definition of a Grassmann algebra}%
\label{ch:DGA}
A {\em Grassmann algebra\/} is a set of formal polynomials of the {\em Grassmann generators\/} $\gamma _{ k}$, $k=1,\cdots,K$. The latter have the properties, ($k,l=1,\cdots,K$)
{\begin{align}{{
 \begin{aligned} 
& \gamma _{ k}\gamma _l = - \gamma _l\gamma _{ k}, & 
& \gamma _{ k}^2 = 0 , \\ 
& \gamma _{ k}\hat B = \hat B\gamma _{ k}, & 
& \gamma _{ k}\hat F = -\hat F\gamma _{ k} , 
\end{aligned}}}%
\label{eq:98RE} 
\end{align}}%
where $\hat B$ and $\hat F$ stand for arbitrary bosonic and fermionic operators. A member of the algebra is a formal polynomial, 
{\begin{multline}\hspace{0.4\columnwidth}\hspace{-0.4\twocolumnwidth} 
g^{(0)} + 
\sum_{1\leq k\leq K}g^{(1)}_{ k} \gamma _{ k} + 
\sum_{1\leq k<l\leq K}g^{(2)}_{kl} \gamma _{ k} \gamma _l + \cdots 
\\ 
+ g^{(K)}_{12\cdots K}\gamma _1\gamma _2\cdots\gamma _K . 
\hspace{0.4\columnwidth}\hspace{-0.4\twocolumnwidth} 
\label{eq:80QL} 
\end{multline}}%
In physical applications, coefficients $g^{(L)}_{k_1k_2\cdots k_L}$, $L=0,\cdots,K$ (coefficient functions, for brevity) may be q-numbers as well as c-numbers. Addition of Grassmann polynomials is addition of coefficient functions, while multiplication is defined as a formal application of the distributive law preserving the order of factors. 
Consistency of formal techniques is warranted by {\em generalised superselection rules\/} which prohibit addition of {\em bosonic\/} and {\em fermionic\/} quantities. The former (latter) is a linear combination of products of even (odd) number of fermionic fields and Grassmann variables. C-numbers and bosonic fields do not count. 

Coefficients in (\ref{eq:80QL}) may depend on parameters such as space-time variables and indices (4-vector, spinor, etc.). So, {\em auxiliary Grassmann variables\/}, or {\em Grassmann fields\/}, are linear combinations of the generators where coefficients are c-number functions \cite{APIII}, 
{\begin{align}{{
 \begin{aligned} 
\psi (x,t) = \sum_{k=1}^{K}\psi _{ k}(x,t)\gamma _{ k} . 
\end{aligned}}}%
\label{eq:81QM} 
\end{align}}%
Variable $x$ has the same meaning as for the channel operators (\ref{eq:57KC}). Grassmann fields are fermionic. They occur only in the intermediate algebra and never turn up in observable quantitities. Properties (\ref{eq:98RE}) extend to Grassmann fields, which commute with bosonic quantities ($B$) and anticommute with fermionic ones ($F$): 
{\begin{align}{{
 \begin{aligned} 
& \psi (x,t)B=B\psi (x,t), & 
& \psi (x,t)F=-F\psi (x,t). 
\end{aligned}}}%
\label{eq:1RH} 
\end{align}}%
\subsection{The uniqueness theorem}%
\label{ch:TUT}
Let $g(x,t)$ be a member of the Grassmann algebra where the coefficients are functions of $x,t$, 
{\begin{align}{{
 \begin{aligned} 
g(x,t) = \sum_{L=0}^K\sum_{\ensuremath{\{
k
\}} }g^{(L)}_{k_1k_2\cdots k_L}(x,t)\gamma _{k_1}\cdots\gamma _{k_L}. 
\end{aligned}}}%
\label{eq:82QN} 
\end{align}}%
To have the derivative by $\psi (x,t)$ uniquely defined we need the following uniqueness theorem. If 
{\begin{align}{{
 \begin{aligned} 
\int dx dt g(x,t)\delta \psi (x,t) = 0
\end{aligned}}}%
\label{eq:83QP} 
\end{align}}%
for an arbitrary Grassmann field $\delta \psi (x,t)$, then $g(x,t)=0$. This theorem does not hold so far as we keep dimensionality $K$ fixed. The obvious (and only) counterexample is 
{\begin{align}{{
 \begin{aligned} 
g(x,t) = g^{(K)}_{12\cdots K}(x,t)\gamma _{1}\cdots\gamma _{K} . 
\end{aligned}}}%
\label{eq:84QQ} 
\end{align}}%
We therefore make $K$ itself a variable. The limit $K\to\infty$ is not implied; $K$ may vary but always stays finite. This spares us all questions of convergence. A general member of the algebra is then specified by (mathematically, {\em is\/}) a countable sequence of coefficient functions, by definition independent of $K$. A Grassmann field is an infinite vector of c-number functions, also independent of $K$. The uniqueness theorem holds if ``arbitrary $\delta \psi (x,t)$'' includes arbitrary coefficient functions {\em and\/} arbitrary $K$. To show that the coefficient function $g^{(L)}_{k_1k_2\cdots k_L}(x,t)$ is zero, it suffices to choose, 
{\begin{align}{{
 \begin{aligned} 
\delta \psi (x,t) = \delta \psi _{L+1}(x,t)\gamma _{L+1}. 
\end{aligned}}}%
\label{eq:85QR} 
\end{align}}%
By construction, \mbox{$k_1<\dots<k_L\leq L$}, and the monomial \mbox{$\gamma _{k_1}\cdots\gamma _{k_L}\gamma _{L+1}$} is nonzero. Hence the coefficient at it in (\ref{eq:83QP}) must be zero: 
{\begin{align}{{
 \begin{aligned} 
\int dx dt g^{(L)}_{k_1k_2\cdots k_L}(x,t)\delta \psi _{L+1}(x,t) = 0. 
\end{aligned}}}%
\label{eq:86QS} 
\end{align}}%
This holds with an arbitrary $\delta \psi _{L+1}(x,t)$, therefore, 
{\begin{align}{{
 \begin{aligned} 
g^{(L)}_{k_1k_2\cdots k_L}(x,t) = 0. 
\end{aligned}}}%
\label{eq:87QT} 
\end{align}}%
\subsection{Grassmann derivative as a variational derivative}%
\label{ch:GDV}
With the uniqueness theorem proved, derivatives by Grassmann fields may be introduced as variational derivatives, 
{\begin{align}{{
 \begin{aligned} 
\delta \mathcal{F}\ensuremath{(
\sigma 
 )} = \int dxdt\hspace{1\auxl}\delta \sigma (x,t)\frac{\delta \mathcal{F}\ensuremath{(\sigma )}}{\delta \sigma (x,t)}
. 
\end{aligned}}}%
\label{eq:88QU} 
\end{align}}%
Dependence of $\mathcal{F}\ensuremath{(
\psi 
 )}$ on $\psi (x,t)$ is understood as dependence of the coefficient functions of the former on those of the latter. This is a generalisation compared to \mbox{Ref.\ \cite{APIII}} where functionals of Grassmann variables were confined to formal Taylor series. Coefficient functions of $\delta \mathcal{F}\ensuremath{(
\psi 
 )}$ are by definition variations of the coefficient functions of $\mathcal{F}\ensuremath{(
\psi 
 )}$. The same applies to $\delta \psi (x,t)$. 

If the variational derivative exists, it is unique; an example of a non-differentiable functional of $\psi (x,t)$ is $\psi _1(x,t)\gamma _2$. Straight from the definition we can prove the relations, 
{\begin{align}{{
 \begin{aligned} 
& \frac{\delta \mathcal{C}}{\delta \psi (x,t)} = 0 , & 
& \frac{\delta \psi (x,t)}{\delta \psi (x',t)} = \delta (x-x')\delta (t-t') , 
\end{aligned}}}%
\label{eq:6RN} 
\end{align}}%
where $\mathcal{C}$ does not depend on $\psi (x)$ (which means that the coefficient functions of the former are independent of those of the latter). By comparing coefficient functions on both sides it is easy to show that, for any two differentiable functionals $\mathcal{F}_1\ensuremath{(\psi )}$ and $\mathcal{F}_2\ensuremath{(\psi )}$, 
{\begin{align}{{
 \begin{aligned} 
& \delta \ensuremath{\big[
\mathcal{F}_1\ensuremath{(\psi )}+\mathcal{F}_2\ensuremath{(\psi )}
\big]} = \delta \mathcal{F}_1\ensuremath{(\psi )}+\delta \mathcal{F}_2\ensuremath{(\psi )}, \\ 
& \delta \ensuremath{\big[
\mathcal{F}_1\ensuremath{(\psi )}\mathcal{F}_2\ensuremath{(\psi )}
\big]} = \delta \mathcal{F}_1\ensuremath{(\psi )}\mathcal{F}_2\ensuremath{(\psi )}+\mathcal{F}_1\ensuremath{(\psi )}\delta \mathcal{F}_2\ensuremath{(\psi )} . 
\end{aligned}}}%
\label{eq:89QV} 
\end{align}}%
From these relations we find the laws of sum and product differentiation, 
{\begin{align}{{
 \begin{aligned} 
& \frac{\delta \ensuremath{\big[
\mathcal{F}_1\ensuremath{(\psi )}+\mathcal{F}_2\ensuremath{(\psi )}
\big]}}{\delta \psi (x,t)}
= \frac{\delta 
\mathcal{F}_1\ensuremath{(\psi )}
}{\delta \psi (x,t)}
+ \frac{\delta 
\mathcal{F}_2\ensuremath{(\psi )}
}{\delta \psi (x,t)}, \\ 
& \frac{\delta \ensuremath{\big[
\mathcal{F}_1\ensuremath{(\psi )}\mathcal{F}_2\ensuremath{(\psi )}
\big]}}{\delta \psi (x,t)}
= \frac{\delta 
\mathcal{F}_1\ensuremath{(\psi )}
}{\delta \psi (x,t)}\mathcal{F}_2\ensuremath{(\psi )}
+ \eta _1\mathcal{F}_1\ensuremath{(\psi )}\frac{\delta 
\mathcal{F}_2\ensuremath{(\psi )}
}{\delta \psi (x,t)}, 
\end{aligned}}}%
\label{eq:91QX} 
\end{align}}%
where $\eta _1=1$ if $\mathcal{F}_1\ensuremath{(\psi )}$ is bosonic and $\eta _1=-1$ if fermionic. Equations (\ref{eq:90QW}) in the main body of the paper partly copy, and partly follow from, eqs.\ (\ref{eq:6RN})--(\ref{eq:91QX}). 
\subsection{Transformation of variational derivatives}%
\label{ch:TAD}
Assume now that we change variables in a functional, 
{\begin{align}{{
 \begin{aligned} 
\mathcal{F}\ensuremath{\big(\psi \big)} = 
\mathcal{G}\ensuremath{\big(\varphi \big)} , 
\end{aligned}}}%
\label{eq:92QY} 
\end{align}}%
where 
{\begin{align}{{
 \begin{aligned} 
\varphi (x,t) = \int dx' dt' K(x,t,x',t')\psi (x',t') . 
\end{aligned}}}%
\label{eq:93QZ} 
\end{align}}%
$K(x,t,x',t')$ is an arbitrary c-number kernel. Equation (\ref{eq:93QZ}) directly extends to variations, 
{\begin{align}{{
 \begin{aligned} 
\delta \varphi (x,t) = \int dx' dt' K(x,t,x',t')\delta \psi (x',t') . 
\end{aligned}}}%
\label{eq:94RA} 
\end{align}}%
We assume that $\mathcal{G}\ensuremath{\big(\varphi \big)}$ is differentiable. Varying the RHS of (\ref{eq:92QY}) we get, 
{\begin{multline}\hspace{0.4\columnwidth}\hspace{-0.4\twocolumnwidth} 
\delta \mathcal{G}\ensuremath{\big(\varphi \big)} 
= \int dx dt \delta \varphi (x,t)
\frac{\delta \mathcal{G}\ensuremath{\big(\varphi \big)}}{\delta \varphi (x,t)}
\\ 
= \int dx dt dx' dt' K(x,t,x',t')\delta \psi (x',t')
\frac{\delta \mathcal{G}\ensuremath{\big(\varphi \big)}}{\delta \varphi (x,t)} . 
\hspace{0.4\columnwidth}\hspace{-0.4\twocolumnwidth} 
\label{eq:95RB} 
\end{multline}}%
This proves differentiability of $\mathcal{F}\ensuremath{\big(\psi \big)}$ and the chain rule, 
{\begin{multline}\hspace{0.4\columnwidth}\hspace{-0.4\twocolumnwidth} 
\frac{\delta \mathcal{F}\ensuremath{\big(\psi \big)}}{\delta \psi (x,t)} 
= \int dx' dt' K(x',t',x,t)
\frac{\delta \mathcal{G}\ensuremath{\big(\varphi \big)}}{\delta \varphi (x',t')} \\ 
= \int dx' dt' \frac{\delta \varphi (x',t')}{\delta \psi (x,t)}
\frac{\delta \mathcal{G}\ensuremath{\big(\varphi \big)}}{\delta \varphi (x',t')} . 
\hspace{0.4\columnwidth}\hspace{-0.4\twocolumnwidth} 
\label{eq:96RC} 
\end{multline}}%
The relation 
{\begin{align}{{
 \begin{aligned} 
\frac{\delta \varphi (x',t')}{\delta \psi (x,t)} = K(x',t',x,t), 
\end{aligned}}}%
\label{eq:97RD} 
\end{align}}%
is simply another form of (\ref{eq:94RA}). Under linear substitutions with c-number coefficients, variational Grassmann derivatives behave in a manner indistinguishable from conventional variational derivatives. 

It should not be forgotten that hidden behind apparent simplicity of eqs.\ (\ref{eq:95RB}), (\ref{eq:96RC}) are infinite sequences of relations for coefficient functions of the Grassmann polynomials. The whole derivation hinges on the uniqueness theorem, which only holds with varying dimensionality of the underlying Grassmann algebra. 
\section{Hori's form of Wick's theorem}%
\label{ch:B}
\subsection{Definitions}%
\label{ch:BD}
Let $K(\tau,\tau')$ be a c-number kernel and $\psi (\tau),{\tilde\psi}(\tau)$ a pair of either c-number or Grassmann fields. We wish to understand the action of the differential operator, 
{\begin{align}{{
 \begin{aligned} 
\mathcal{K} & = \int d\tau d\tau'K(\tau,\tau')\frac{\delta }{\delta \psi (\tau)}
\frac{\delta }{\delta {\tilde\psi}(\tau') }, 
\end{aligned}}}%
\label{eq:17MS} 
\end{align}}%
on multitime products of the fields. 
We also introduce the notation, 
{\begin{align} 
\label{eq:18MT} 
\Psi(\tau) & = \int d\tau' K(\tau,\tau')
\frac{\delta }{\delta {\tilde\psi}(\tau') }, \\ 
\tilde\Psi (\tau) & = \int d\tau' K(\tau',t)\frac{\delta }{\delta \psi (\tau')} . 
\label{eq:15MQ} 
\end{align}}%
Variable $\tau $ is thought about as time, but may in fact have any meaning, so far as the integration $\int d\tau $ and the symmetric delta-function $\delta (\tau,\tau')$ remain defined. This implies the consistency conditions, 
{\begin{align}{{
 \begin{aligned} 
& \int d\tau' \delta (\tau,\tau') \mathcal{X}(\tau') = \mathcal{X}(\tau), \\ 
& \frac{\delta }{\delta \psi (\tau)}\psi (\tau') = 
\frac{\delta }{\delta \tilde\psi (\tau)}\tilde\psi (\tau') = \delta (\tau,\tau'), \\
& \frac{\delta }{\delta \psi (\tau)}\tilde\psi (\tau') = 
\frac{\delta }{\delta \tilde\psi (\tau)}\psi (\tau') = 0. 
\end{aligned}}}%
\label{eq:48PA} 
\end{align}}%
where $\mathcal{X}(\tau)$ may be a c-number field, Grassmann field, or the corresponding functional derivatives. Obviously, these must be defined as well, with the corresponding commutation/anticommutation properties. These mathematical requirements are not restrictive, and the formal pattern of field products with contractions characteristic of Wick's theorem may be generalised pretty far. The deterring factor is ones ability to interpret the kernel $K(\tau,\tau')$ as a contraction of a pair of operators, cf.\ eq.\ (\ref{eq:38NQ}) below. Without it, all such generalisations are meaningless (at least, for physics). One example when such generalisation makes perfect sense is the C-contour.

\subsection{Intermediate terms}%
\label{ch:BI}
Using (\ref{eq:48PA}) it is straighforward to verify the formulae, 
{\begin{align}{{
 \begin{aligned} 
\mathcal{K}\psi (\tau)\mathcal{P} & = \psi (\tau)\mathcal{K}\mathcal{P} +\varepsilon_{\mathrm{f}} \Psi (\tau)\mathcal{P} , \\ 
\mathcal{K}{\tilde\psi} (\tau)\mathcal{P} & = {\tilde\psi} (\tau)\mathcal{K}\mathcal{P} + \tilde\Psi (\tau)\mathcal{P} , 
\end{aligned}}}%
\label{eq:19MU} 
\end{align}}%
where $\mathcal{P}$ stands for an arbitrary product of the fields. The sign factor $\varepsilon_{\mathrm{f}}$ equals 1 when \mbox{$
\psi (\tau ),{\tilde\psi}(\tau )
$} are c-numbers and $-1$ for Grassmann variables. Now, let $\mathcal{P}_n$ be a product of $n$ factors. Applying (\ref{eq:19MU}) recursively we see that $\mathcal{K}\mathcal{P}_n$ is a sum of $n$ terms, such that in each term one incidence of lowercase $\psi $ is replaced by uppercase $\Psi $, 
{\begin{align}{{
 \begin{aligned} 
& \psi (\tau) \to \varepsilon_{\mathrm{f}} \Psi (\tau), & 
& \tilde\psi (\tau) \to \tilde\Psi (\tau)
\end{aligned}}}%
\label{eq:34NL} 
\end{align}}%
For example, 
{\begin{align}{{
 \begin{aligned} 
& \mathcal{K} \psi (\tau)\tilde\psi(\tau') 
= \varepsilon_{\mathrm{f}}\Psi (\tau)\tilde\psi(\tau') , \\ 
& \mathcal{K} \tilde\psi(\tau')\psi (\tau)
= \tilde\Psi(\tau')\psi (\tau) , \\ 
& {\begin{aligned}\mathcal{K} \tilde\psi(\tau')\psi (\tau)\tilde\psi(\tau'')
& = \tilde\Psi(\tau')\psi (\tau)\tilde\psi(\tau'') 
\\ & \ \ \ 
+\varepsilon_{\mathrm{f}}\tilde\psi(\tau')\Psi (\tau)\tilde\psi(\tau'') . \end{aligned}}
\end{aligned}}}%
\label{eq:32NJ} 
\end{align}}%
Terms with $\tilde\Psi (\tau)$ or $\Psi (\tau)$ in the rightmost position are zero and have been omitted. No coefficients occur, except that in the fermionic case terms with $\Psi (\tau)$ acquire a minus. Similarly, $\mathcal{K}^2$ turns two factors uppercase, e.g., 
{\begin{multline} 
\mathcal{K}^2 
\tilde\psi(\tau)\psi (\tau')\psi (\tau'')\tilde\psi(\tau''')
= \mathcal{K}\ensuremath{\big[
\tilde\Psi(\tau)\psi (\tau')\psi (\tau'')\tilde\psi(\tau''') \\ 
+\varepsilon_{\mathrm{f}}\tilde\psi(\tau)\Psi (\tau')\psi (\tau'')\tilde\psi(\tau''')
+\varepsilon_{\mathrm{f}}\tilde\psi(\tau)\psi (\tau')\Psi (\tau'')\tilde\psi(\tau''')
\big]} \\ 
=
2\ensuremath{\big[
\varepsilon_{\mathrm{f}} \tilde\Psi(\tau)\Psi (\tau')\psi (\tau'')\tilde\psi(\tau''') 
+\varepsilon_{\mathrm{f}} \tilde\Psi(\tau)\psi (\tau')\Psi (\tau'')\tilde\psi(\tau''') \\ 
+ \tilde\psi(\tau)\Psi (\tau')\Psi (\tau'')\tilde\psi(\tau''')
\big]} 
. 
\label{eq:33NK} 
\end{multline}}%
Terms with $\tilde\Psi (\tau)$ or $\Psi (\tau)$ in the rightmost position are again dropped. In obtaining (\ref{eq:33NK}) we also used 
the commutativity properties, 
{\begin{align}{{
 \begin{aligned} 
& \mathcal{K}\Psi (\tau) = \Psi (\tau)\mathcal{K}, & 
& \mathcal{K}\tilde\Psi (\tau) = \tilde\Psi (\tau)\mathcal{K}
\end{aligned}}}%
\label{eq:36NN} 
\end{align}}%
Each term in (\ref{eq:33NK}) occurs twice, depending on the order of replacements. Signs of terms do not depend on this order, so that identical terms always add up. It is clear that these rules will also work for $\mathcal{K}^m$ applied to an arbitrary product of $n$ factors. For $m>n$, the result is zero, otherwise it is a sum of terms, obtained by all possible sets of $m$ replacements (\ref{eq:34NL}), with signs determined only by the latter, with the overall coefficient $m!$. In the below we call them {\em intermediate terms\/}. 

\subsection{Wick terms}%
\label{ch:BW}
Recall now that $\tilde\Psi (\tau)$ are $\Psi (\tau)$ are differential operators. It is straigtforward to verify the recursion formulae, 
{\begin{align}{{
 \begin{aligned} 
\Psi (\tau)\psi (\tau')\mathcal{P} 
& = \varepsilon_{\mathrm{f}}\psi (\tau')\Psi (\tau)\mathcal{P}, \\ 
\tilde\Psi (\tau)\tilde\psi (\tau')\mathcal{P} 
& = \varepsilon_{\mathrm{f}}\tilde\psi (\tau')\tilde\Psi (\tau)\mathcal{P} , 
\end{aligned}}}%
\label{eq:16MR} 
\end{align}}%
and 
{\begin{align}{{
 \begin{aligned} 
\Psi (\tau)\tilde\psi (\tau')\mathcal{P} 
& = \varepsilon_{\mathrm{f}}\tilde\psi (\tau')\Psi (\tau)\mathcal{P} + K(\tau,\tau')\mathcal{P}, \\ 
\tilde\Psi (\tau)\psi (\tau')\mathcal{P} 
& = \varepsilon_{\mathrm{f}}\psi (\tau')\tilde\Psi (\tau)\mathcal{P} + K(\tau',t)\mathcal{P} . 
\end{aligned}}}%
\label{eq:37NP} 
\end{align}}%
To visualise these relations, think about uppercase $\Psi $'s skipping along the product of lowercase $\psi$'s. The pairs $\Psi (\tau)\psi (\tau')$ and $\tilde\Psi (\tau)\tilde\psi (\tau')$ have no choice: they must (anti)commute. The pairs $\Psi (\tau)\tilde\psi (\tau')$ and $\tilde\Psi (\tau)\psi (\tau')$ have a choice: they can either (anti)commute or ``contract.'' Contracted pairs cancel, leaving instead c-number kernels $K(\tau,\tau')$. Terms where an uppercase $\Psi $ has reached the rightmost position are zero, so that $m$ uppercase $\Psi $'s either give rise to $m$ contractions or ``perish.'' (At this stage, we use the term ``contraction'' rather loosely. Its exact meaning is specified shortly.) Each of intermediate terms created by application of $\mathcal{K}^m$ to a product of fields thus results in a sum of typical {\em Wick terms\/} with $m$ field pairs replaced by contractions. It is easy to see that any given Wick term is ``parented'' by a unique intermediate term (specified by left ends of contractions), and that all Wick terms ``parented'' by a particular intermediate term differ (distinguished by right ends of contractions). Hence, up to the overall coefficient of $m!$ (and signs, see below), the differential operator $\mathcal{K}^m$ applied to a product of fields generates a sum of all possible Wick terms with $m$ contractions, each occuring only once. 

\subsection{Sign rules}%
\label{ch:BS}
Consider now signs of contractions. By itself, the act of contraction introduces no sign, cf.\ eqs.\ (\ref{eq:37NP}) where the terms with c-number kernels occur with a plus. Sign factors due to replacements (\ref{eq:34NL}) are absorbed by definitions of contractions, (in Wick's original dot notation \cite{Wick})
{\begin{align}{{
 \begin{aligned} 
{\hat\psi}{}^{{\mbox{\rm\boldmath$\cdot$}}}(\tau) {\hat{\tilde\psi}}{}^{{\mbox{\rm\boldmath$\cdot$}}}(\tau') 
& =\varepsilon_{\mathrm{f}}\Psi(\tau){\tilde\psi}(\tau')= \varepsilon_{\mathrm{f}} K(\tau,\tau'), \\ 
{\hat{\tilde\psi}}{}^{{\mbox{\rm\boldmath$\cdot$}}}(\tau') {\hat\psi}{}^{{\mbox{\rm\boldmath$\cdot$}}}(\tau) 
& = \tilde\Psi(\tau')\psi (\tau)= K(\tau,\tau'). 
\end{aligned}}}%
\label{eq:38NQ} 
\end{align}}%
These relations are consistent with the property, 
{\begin{align}{{
 \begin{aligned} 
{\hat\psi}{}^{{\mbox{\rm\boldmath$\cdot$}}}(\tau) {\hat{\tilde\psi}}{}^{{\mbox{\rm\boldmath$\cdot$}}}(\tau') 
= \varepsilon_{\mathrm{f}} {\hat{\tilde\psi}}{}^{{\mbox{\rm\boldmath$\cdot$}}}(\tau') {\hat\psi}{}^{{\mbox{\rm\boldmath$\cdot$}}}(\tau), 
\end{aligned}}}%
\label{eq:39NR} 
\end{align}}%
which must hold for contractions in the true meaning of the term. 

The only nontrivial source of signs are therefore eqs.\ (\ref{eq:16MR}), (\ref{eq:37NP}). Sign of any Wick term may be worked out by formally commuting uppercase $\Psi $'s to the left of lowercase $\psi $ with which they are contracted. It is obvious that this rule coincides with the sign rule of Wick's theorem, which requires contracted fermionic pairs to be formally commuted to adjacent positions. 

\subsection{The reordering exponent}%
\label{ch:BE}
This way, the differential operator $\mathcal{K}^m$ applied to an arbitrary product of fields generates a correct pattern of terms with $m$ contractions with the overall factor of $m!$. In the fermionic case, anticommutativity of Grassmann derivatives assures correct signs of all terms. 
The total differential operator to be used in Hori's form of Wick's theorem is therefore found to be, 
{\begin{align}{{
 \begin{aligned} 
1+\sum_{m=1}^{\infty}\frac{\mathcal{K}^m}{m!}= \exp\mathcal{K} . 
\end{aligned}}}%
\label{eq:22MX} 
\end{align}}%
Vital for consistency of this interpretation are eqs.\ (\ref{eq:38NQ}), (\ref{eq:39NR}), which allowed us to redefine the arbitrary kernel $K(\tau,\tau')$ as a genuine contraction of a pair of operators. 

\subsection{Real fields}%
\label{ch:BR}
This reasoning is easily adjusted to the real case where $\psi (\tau )={\tilde\psi}(\tau )$. We allow $\psi (\tau )$ to be a Grassmann variable, thus extending the Hori-Wick theorem to (hypothetical) Hermitian fermionic fields. We also assume that 
{\begin{align}{{
 \begin{aligned} 
K(\tau ,\tau ') = \varepsilon_{\mathrm{f}} K(\tau ',\tau ) . 
\end{aligned}}}%
\label{eq:93VH} 
\end{align}}%
The $\mathcal{K}$ form and the uppercase $\Psi (\tau )$ are defined as, 
{\begin{align}{{
 \begin{aligned} 
& \mathcal{K} = \frac{1}{2}\int d\tau d\tau'K(\tau,\tau')
\frac{\delta }{\delta \psi (\tau)}
\frac{\delta }{\delta \psi (\tau') }, \\ 
& \Psi (\tau) = \int d\tau' K(\tau',t)\frac{\delta }{\delta \psi (\tau')} . 
\end{aligned}}}%
\label{eq:90VD} 
\end{align}}%
The rules of commutation are, 
{\begin{align}{{
 \begin{aligned} 
\mathcal{K}\Psi (\tau)\mathcal{P} & = \Psi (\tau)\mathcal{K}\mathcal{P}, \\ 
\mathcal{K}\psi (\tau)\mathcal{P} & = \psi (\tau)\mathcal{K}\mathcal{P} + \Psi (\tau)\mathcal{P} , \\ 
\Psi (\tau)\psi (\tau')\mathcal{P} 
& = \varepsilon_{\mathrm{f}}\psi (\tau')\Psi (\tau)\mathcal{P} + K(\tau',t)\mathcal{P} . 
\end{aligned}}}%
\label{eq:91VE} 
\end{align}}%
Connection to physics is established by the definition, 
{\begin{align}{{
 \begin{aligned} 
{\hat\psi}{}^{{\mbox{\rm\boldmath$\cdot$}}}(\tau ){\hat\psi}{}^{{\mbox{\rm\boldmath$\cdot$}}}(\tau' ) = K(\tau ',\tau ) = \varepsilon_{\mathrm{f}} K(\tau, \tau' ) . 
\end{aligned}}}%
\label{eq:92VF} 
\end{align}}%
The only source of signs is then the last of eqs.\ (\ref{eq:91VE}). The rest of reasoning generalises trivially.
\section{Algebra of Green functions of a quantum channel}%
\label{ch:A}
\subsection{The commutator and related quantities}%
\label{ch:RAC}
Up to the fact that $\tilde \Delta _{\text{R}}$ was defined with transposition of arguments, the algebra of Green functions of a quantum channel is exactly that of a conventional complex field. We therefore only list formulae with terse comments. For more details see our papers \cite{API,APIII}. 
For conventional quantised fields, many of the relations in this appendix appear in various textbooks on quantum field theory. Some may be found in Schwinger's series on quantum field theory \cite{SchwingerR,SchwingerS}. 

All kernels we introduce are related to the commutator, 
{\begin{align}{{
 \begin{aligned} 
-i\hbar\Delta(x,x',t-t') = \ensuremath{\big\langle 0\big|
\ensuremath{\big[
{\hat\psi} (x,t),{\hat{\tilde\psi}}(x',t')
\big]} _{-\varepsilon_{\mathrm{f}}} 
\big|0\big\rangle} , 
\end{aligned}}}%
\label{eq:70KS} 
\end{align}}%
where we have extended the notation 
with the sign factor to the commutator, 
{\begin{align}{{
 \begin{aligned} \ensuremath{\big[
\hat A,\hat B
\big]} _{-\varepsilon_{\mathrm{f}}} = \hat A \hat B -\varepsilon_{\mathrm{f}}\hat B\hat A . 
\end{aligned}}}%
\label{eq:63FL} 
\end{align}}%
We remind that ``commutator'' means commutator for bosons and anticommutator for fermions.
Quantity (\ref{eq:70KS}) is a c-number, so putting it under the averaging is a matter of unification of all definitions. Subtracting the last of eqs.\ (\ref{eq:90LP}) from the next-to-the-last one we find, 
{\begin{multline}\hspace{0.4\columnwidth}\hspace{-0.4\twocolumnwidth} 
\Delta(x,x',t-t') =\Delta^{(+)} (x,x',t-t')+\Delta^{(-)} (x,x',t-t'). 
\hspace{0.4\columnwidth}\hspace{-0.4\twocolumnwidth} 
\label{eq:71KT} 
\end{multline}}%
Unlike (\ref{eq:70KS}), in (\ref{eq:90LP}) vacuum averaging is essential. 
Using (\ref{eq:57KC}), 
{\begin{align}{{
 \begin{aligned} 
\Delta^{(+)} (x,x',t-t') 
& = i\sum_{\kappa}\frac{ u_{\kappa}(x){\tilde u}_{\kappa}(x')\textrm{e}^{-i\omega _{\kappa}(t-t')}} 
{2\omega _{\kappa }}, \\ 
\Delta^{(-)} (x,x',t-t') 
& = -i\varepsilon_{\mathrm{f}}\sum_{\kappa}\frac{ v_{\kappa}(x'){\tilde v}_{\kappa}(x)\textrm{e}^{i\omega _{\kappa}(t-t')}} 
{2\omega _{\kappa }} . 
\end{aligned}}}%
\label{eq:73KV} 
\end{align}}%
We see that $\Delta^{(+)} $ is purely frequency-positive\ and $\Delta^{(-)} $ is purely frequency-negative. Notation $\Delta ^{(\pm)}$ is consistent with these kernels being the frequency-positive and negative\ parts of the commutator. This property will be instrumental in expressing contractions by the the transfer functions. 
\subsection{Response functions}%
\label{ch:RAR}
In terms of $\Delta $, 
eqs.\ (\ref{eq:76KY}) and (\ref{eq:78LA}) become, 
{\begin{align} 
\label{eq:94LT} 
\Delta_{\text{R}}(x,x',t-t') & = \theta(t-t')\Delta (x,x',t-t'), \\
\tilde\Delta_{\text{R}}(x,x',t-t') & = -\theta(t'-t)\Delta (x,x',t-t').
\label{eq:89LN} 
\end{align}}%
Explicit expressions for the transfer functions 
are found using eqs.\ (\ref{eq:73KV}). 
\subsection{Contractions}%
\label{ch:Co}
The kernels $\Delta ^{(\pm)}$ occur as contractions in Wick's theorem, cf.\ eqs.\ (\ref{eq:90LP}). The other two contractions are expressed by them, 
{\begin{align} 
\Delta _{\text{F}}(x,x',t-t') 
& = \theta(t-t')\Delta ^{(+)}(x,x',t-t')
\nonumber 
\\ & \ \ \ - \theta(t'-t)\Delta ^{(-)}(x,x',t-t'), 
\label{eq:96LV} 
\\
\tilde\Delta _{\text{F}}(x,x',t-t') 
& = \theta(t-t')\Delta ^{(-)}(x,x',t-t')
\nonumber 
\\ & \ \ \ - \theta(t'-t)\Delta ^{(+)}(x,x',t-t') . 
\label{eq:92LR} 
\end{align}}%
Explicit expressions again follow from (\ref{eq:73KV}). 
\subsection{Response transformation of the quantum Green functions}%
\label{ch:RT}
We now express contractions (\ref{eq:90LP}) by the transfer functions (\ref{eq:14RW}). We follow \mbox{Ref.\ \cite{API}}, the case of free complex fields, with two inessential amendments: there are now two formally independent transfer functions, and transposition of arguments occuring in response transformations in \mbox{Ref.\ \cite{API}} is now built into the definition of $\tilde\Delta _{\text{R}}$. 

Adding $\Delta ^{(-)}$ to (\ref{eq:96LV}) and $\Delta ^{(+)}$ to (\ref{eq:92LR}) and recalling (\ref{eq:71KT}), 
(\ref{eq:94LT}) and (\ref{eq:89LN}) we find, 
{\begin{align}{{
 \begin{aligned} 
\Delta _{\text{F}}(x,x',t-t') +\Delta ^{(-)}(x,x',t-t') 
=\Delta_{\text{R}}(x,x',t-t') , \\ 
\tilde\Delta _{\text{F}}(x,x',t-t') +\Delta ^{(+)}(x,x',t-t') 
=\tilde\Delta_{\text{R}}(x,x',t-t') . 
\end{aligned}}}%
\label{eq:97LW} 
\end{align}}%
Subtracting (\ref{eq:89LN}) from (\ref{eq:94LT}) yields, 
{\begin{align}{{
 \begin{aligned} 
\Delta_{\text{R}}(x,x',t-t') -\tilde\Delta_{\text{R}}(x,x',t-t') 
=\Delta(x,x',t-t'). 
\end{aligned}}}%
\label{eq:93LS} 
\end{align}}%
Separating the frequency-positive and negative\ parts in this relation results in the first pair of eqs.\ (\ref{eq:60WS}). 
Substituting them into eqs.\ (\ref{eq:97LW}) and recalling (\ref{eq:82RZ}) we recover the second pair of eqs.\ (\ref{eq:60WS}). Up to the aforementioned two distinctions, response transformations for the quantum channel are indeed identical to those for a complex field \cite{API,APIII}. 
\section{Derivation of equations (\ref{eq:76KY})--(\ref{eq:22SE})}%
\label{ch:DER}
In the first order in $\hat H_{\text{I}}$, the Heisenberg channel operators are given by standard expressions, 
{\begin{align} 
\label{eq:79LB} 
{\hat\Psi}(x,t) = {\hat\psi}(x,t) - \frac{i}{\hbar }\int_{-\infty}^{t} dt'
\int dx'\ensuremath{\big[
{\hat\psi}(x,t),\hat H_{\text{I}}(x',t')
\big]} , \\ 
{\hat{\tilde\Psi}}(x,t) = {\hat{\tilde\psi}}(x,t) - \frac{i}{\hbar }\int_{-\infty}^{t} dt'
\int dx'\ensuremath{\big[
{\hat{\tilde\psi}}(x,t),\hat H_{\text{I}}(x',t')
\big]} . 
\label{eq:15RX} 
\end{align}}%
Consider firstly the commutator in eq.\ (\ref{eq:79LB}), 
{\begin{multline}\hspace{0.4\columnwidth}\hspace{-0.4\twocolumnwidth} 
\ensuremath{\big[
{\hat\psi}(x,t),\hat H_{\text{I}}(x',t')
\big]} \\ 
= {\hat\psi}{\hat{\tilde\psi}}{}'\sigma '+{\hat\psi}\tilde\sigma '{\hat\psi}' 
- {\hat{\tilde\psi}}{}'\sigma'{\hat\psi}-\tilde\sigma '{\hat\psi}'{\hat\psi} . 
\hspace{0.4\columnwidth}\hspace{-0.4\twocolumnwidth} 
\label{eq:80LC} 
\end{multline}}%
For brevity, we write, ${\hat\psi}={\hat\psi}(x,t)$, ${\hat\psi}'={\hat\psi}(x',t')$, and so on. To apply eqs.\ (\ref{eq:90QW}), $\sigma '$ in all terms must be commuted to the left. With eqs.\ (\ref{eq:13RV}) we have, 
{\begin{align}{{
 \begin{aligned} 
{\hat\psi}{\hat{\tilde\psi}}{}'\sigma ' & = \sigma '{\hat\psi}{\hat{\tilde\psi}}{}', \\ 
{\hat\psi}\tilde\sigma '{\hat\psi}' & = \varepsilon_{\mathrm{f}} \tilde\sigma '{\hat\psi}{\hat\psi}', & 
{\hat{\tilde\psi}}{}'\sigma'{\hat\psi} & = \varepsilon_{\mathrm{f}} \sigma'{\hat{\tilde\psi}}{}'{\hat\psi}. 
\end{aligned}}}%
\label{eq:17RZ} 
\end{align}}%
In the first term, the source commutes with two operators, causing (for fermions) two sign changes which cancel each other. In the other two, it commutes only once; for fermions, the sign change persists, resulting in the commutator turning anticommutator: 
{\begin{align}{{
 \begin{aligned} 
\ensuremath{\big[
{\hat\psi}(x,t),\hat H_{\text{I}}(x',t')
\big]} 
= \sigma '\ensuremath{\big[
{\hat\psi},{\hat{\tilde\psi}}{}'
\big]}_{-\varepsilon_{\mathrm{f}}} -\tilde\sigma '\ensuremath{\big[
{\hat\psi}',{\hat\psi}
\big]}_{-\varepsilon_{\mathrm{f}}} . 
\end{aligned}}}%
\label{eq:18SA} 
\end{align}}%
We use notation (\ref{eq:63FL}). By direct calculation, 
{\begin{align}{{
 \begin{aligned} 
\ensuremath{\big[
{\hat\psi}',{\hat\psi}
\big]}_{-\varepsilon_{\mathrm{f}}} = 0, 
\end{aligned}}}%
\label{eq:19SB} 
\end{align}}%
and we recover eq.\ (\ref{eq:76KY}). Similarly, 
{\begin{align}{{
 \begin{aligned} 
\ensuremath{\big[
{\hat{\tilde\psi}}(x,t),\hat H_{\text{I}}(x',t')
\big]} 
= \sigma '\ensuremath{\big[
{\hat{\tilde\psi}},{\hat{\tilde\psi}}{}'
\big]}_{-\varepsilon_{\mathrm{f}}} -\tilde\sigma '\ensuremath{\big[
{\hat\psi}',{\hat{\tilde\psi}}
\big]}_{-\varepsilon_{\mathrm{f}}} , 
\end{aligned}}}%
\label{eq:20SC} 
\end{align}}%
and 
{\begin{align}{{
 \begin{aligned} 
\ensuremath{\big[
{\hat{\tilde\psi}},{\hat{\tilde\psi}}{}'
\big]}_{-\varepsilon_{\mathrm{f}}} = 0, 
\end{aligned}}}%
\label{eq:21SD} 
\end{align}}%
resulting in eq.\ (\ref{eq:78LA}). Equations (\ref{eq:22SE}) are a consequence of eqs.\ (\ref{eq:19SB}) and (\ref{eq:21SD}). 
\section{Proof of Wick's theorem for the quantum channel}%
\label{ch:PW}
We assume known Wick's theorem for the harmonic oscillator and for the two-level system. Namely, 
{\begin{multline}\hspace{0.4\columnwidth}\hspace{-0.4\twocolumnwidth} 
{\begin{aligned}
& {\begin{aligned}
& T_C\mathcal{F}\ensuremath{\big(
\hat b_{\kappa+},\hat b_{\kappa+}^{\dag},
\hat b_{\kappa-},\hat b_{\kappa-}^{\dag}
 \big)} 
= {\mbox{\rm\boldmath$:$}}\mathcal{F}_N\ensuremath{\big(
\hat b_{\kappa},\hat b_{\kappa}^{\dag}
 \big)}
{\mbox{\rm\boldmath$:$}} , 
& \mathrm{where}\end{aligned}} 
\\ 
&\mathcal{F}_N\ensuremath{\big(
b_{\kappa+},\tilde b_{\kappa+}
 \big)} 
= \exp Z^{(k)}_C\ensuremath{\bigg(
\frac{\delta }{\delta b_{\kappa+}},
\frac{\delta }{\delta \tilde b_{\kappa+}},
\frac{\delta }{\delta b_{\kappa-}},
\frac{\delta }{\delta \tilde b_{\kappa-}}
 \bigg)} 
\end{aligned}} 
\\ \times \mathcal{F}\ensuremath{\big(
b_{\kappa+},\tilde b_{\kappa+},
b_{\kappa-},\tilde b_{\kappa-}
 \big)}\settoheight{\auxlv}{$|$}%
\raisebox{-0.3\auxlv}{$|_{b_{\kappa-}=b_{\kappa+},\tilde b_{\kappa-}=\tilde b_{\kappa+}}$}. 
\hspace{0.4\columnwidth}\hspace{-0.4\twocolumnwidth} 
\label{eq:33SS} 
\end{multline}}%
In this relation, 
{\begin{align}{{
 \begin{aligned} 
& \hat b_{\kappa}(t) = \hat b_{\kappa} \textrm{e}^{-i\omega _{\kappa}t}, & 
& \hat b_{\kappa}^{\dag}(t) = \hat b_{\kappa}^{\dag} \textrm{e}^{i\omega _{\kappa}t} , 
\end{aligned}}}%
\label{eq:34ST} 
\end{align}}%
where $\hat b_{\kappa},\hat b_{\kappa}^{\dag}$ are specified by eq.\ (\ref{eq:58KD}), and $b_{\kappa\pm}(t),\tilde b_{\kappa\pm}(t)$ are four auxiliary functional variables, c-number or Grassmann. The reordering form reads, 
{\begin{multline}\hspace{0.4\columnwidth}\hspace{-0.4\twocolumnwidth} 
Z_C^{(k)}\ensuremath{\big(
{\tilde f_+},
{f_+},
{\tilde f_-},
{f_-}
 \big)} \\ 
= -i\varepsilon_{\mathrm{f}}\int dt dt'\ensuremath{\big[
g_{\textrm{F}\kappa }(t-t')
{\tilde f_+(t)}{f_+(t')}
\\ -g_{\textrm{F}\kappa }^*(t'-t)
{\tilde f_-(t)}{f_-(t')}
+ g_{\kappa}^{(+)}(t-t')
{\tilde f_-(t)}{f_+(t')}
\big]} 
, 
\hspace{0.4\columnwidth}\hspace{-0.4\twocolumnwidth} 
\label{eq:35SU} 
\end{multline}}%
where the kernels are given by the formulae, 
{\begin{align}{{
 \begin{aligned} 
g_{\kappa}^{(+)}(t-t') & = i\textrm{e}^{-i\omega _{\kappa}(t-t')}, \\ 
g_{\textrm{F}k}(t-t') & = i\theta(t-t')\textrm{e}^{-i\omega _{\kappa}(t-t')} , 
\end{aligned}}}%
\label{eq:36SV} 
\end{align}}%
and the sign factor was defined by eq.\ (\ref{eq:60FH}). 
An identical set of relations holds for the antiparticle operators, 
{\begin{align}{{
 \begin{aligned} 
& \hat c_{\kappa}(t) = \hat c_{\kappa} \textrm{e}^{-i\omega _{\kappa}t}, & 
& \hat c_{\kappa}^{\dag}(t) = \hat c_{\kappa}^{\dag} \textrm{e}^{i\omega _{\kappa}t} . 
\end{aligned}}}%
\label{eq:37SW} 
\end{align}}%
For fermions, the series for the exponent is truncated, but this does not affect the ensuing argument. 

Equations (\ref{eq:33SS})--(\ref{eq:35SU}) are eqs.\ (\ref{eq:60PP})--(\ref{eq:90LP}) downgraded to the nonrelativistic\ case for a single mode. Formulae (\ref{eq:36SV}) for contractions may be verified in close similarity to eqs.\ (\ref{eq:90LP}). The difference is that, unlike quantum channel, Wick's theorem for a single mode is a well established fact which we can rely upon. 

An arbitrary functional of the channel operators may be brought to the normal form by applying the Hori-Wick theorem (\ref{eq:33SS}) modewise. The channel operators are then regarded not entities as in eq.\ (\ref{eq:60PP}), but linear combinations of $\hat b_{\kappa}(t),\hat c_{\kappa}(t),\hat b_{\kappa}^{\dag}(t),\hat c_{\kappa}^{\dag}(t)$. The computational part of the Hori-Wick theorem turns into, 
\begin{widetext} 
{\begin{multline} 
\mathcal{F}_N( \psi_+,{\tilde\psi}_+ ) 
= \exp \ensuremath{\bigg\{
\sum_{\kappa}\ensuremath{\bigg[
Z^{(k)}_C\ensuremath{\bigg(
\frac{\delta }{\delta b_{\kappa+}},
\frac{\delta }{\delta \tilde b_{\kappa+}},
\frac{\delta }{\delta b_{\kappa-}},
\frac{\delta }{\delta \tilde b_{\kappa-}}
 \bigg)} 
+Z^{(k)}_C\ensuremath{\bigg(
\frac{\delta }{\delta c_{\kappa+}},
\frac{\delta }{\delta \tilde c_{\kappa+}},
\frac{\delta }{\delta c_{\kappa-}},
\frac{\delta }{\delta \tilde c_{\kappa-}}
 \bigg)}
\bigg]} 
\bigg\}} 
\\ \times 
\mathcal{F}( \psi_+, {\tilde\psi}_+ ,\psi_-, {\tilde\psi}_-) 
\settoheight{\auxlv}{$|$}%
\raisebox{-0.3\auxlv}{$|_{
b_{\kappa-}=b_{\kappa+},\tilde b_{\kappa-}=\tilde b_{\kappa+}, 
c_{\kappa-}=c_{\kappa+},\tilde c_{\kappa-}=\tilde c_{\kappa+}, 
}$}, 
\label{eq:38SX} 
\end{multline}}%
\end{widetext}%
where 
{\begin{align}{{
 \begin{aligned} 
\psi_{\pm} (x,t) & = \sum_{\kappa}
\sqrt{\frac{\hbar }{2\omega_{\kappa }}}
\ensuremath{\big[
 u_{\kappa}(x) \hspace{1\auxl} b_{\kappa\pm}(t)
+\tilde v_{\kappa}(x) \hspace{1\auxl}\tilde c_{\kappa\pm}(t)
\big]} , \\
{\tilde\psi}_{\pm} (x,t) & = \sum_{\kappa} 
\sqrt{\frac{\hbar }{2\omega_{\kappa }}}
\ensuremath{\big[
 v_{\kappa}(x) \hspace{1\auxl} c_{\kappa\pm}(t)
+\tilde u_{\kappa}(x) \hspace{1\auxl}\tilde b_{\kappa\pm}(t)
\big]} . 
\end{aligned}}}%
\label{eq:39SY} 
\end{align}}%
Strictly speaking, eq.\ (\ref{eq:38SX}) holds only if its RHS remains a functional of fields (\ref{eq:39SY}). Indeed, with eq.\ (\ref{eq:39SY}) regarded a change of functional variables, the standard chain rules are, 
{\begin{align}{{
 \begin{aligned} 
\frac{\delta }{\delta b_{\kappa\pm}(t)} 
= \sqrt{\frac{\hbar }{2\omega_{\kappa }}}
\int dx\hspace{1\auxl} u_{\kappa}(x) 
\frac{\delta }{\delta \psi_{\pm}(x,t)}, 
\\ 
\frac{\delta }{\delta c_{\kappa\pm}(t)} 
= \sqrt{\frac{\hbar }{2\omega_{\kappa }}}
\int dx\hspace{1\auxl} v_{\kappa}(x) 
\frac{\delta }{\delta {\tilde\psi}_{\pm}(x,t)}, 
\\ 
\frac{\delta }{\delta\tilde b_{\kappa\pm}(t)} 
= \sqrt{\frac{\hbar }{2\omega_{\kappa }}}
\int dx\hspace{1\auxl} \tilde u_{\kappa}(x) 
\frac{\delta }{\delta \tilde\psi_{\pm}(x,t)}, 
\\ 
\frac{\delta }{\delta\tilde c_{\kappa\pm}(t)} 
= \sqrt{\frac{\hbar }{2\omega_{\kappa }}}
\int dx\hspace{1\auxl} \tilde v_{\kappa}(x) 
\frac{\delta }{\delta \psi _{\pm}(x,t)} . 
\end{aligned}}}%
\label{eq:41TA} 
\end{align}}%
Differentiations in (\ref{eq:38SX}) may therefore be rewritten in terms of entire fields (\ref{eq:39SY}), and the modewise variables $\hat b_{\kappa}(t),\hat c_{\kappa}(t),\hat b_{\kappa}^{\dag}(t),\hat c_{\kappa}^{\dag}(t)$ excluded. 
The key property of variational Grassmann derivatives is that, under substitutions with c-number coefficients, chain rules for them are identical with those for conventional variational derivatives (appendix \ref{ch:TAD}). Chain rules (\ref{eq:41TA}) thus equally apply to bosons and fermions. 
Using these relations it is straightforward to show that, 
{\begin{multline}\hspace{0.4\columnwidth}\hspace{-0.4\twocolumnwidth} 
\sum_{\kappa}\ensuremath{\bigg[
Z^{(k)}_C\ensuremath{\bigg(
\frac{\delta }{\delta b_{\kappa+}},
\frac{\delta }{\delta \tilde b_{\kappa+}},
\frac{\delta }{\delta b_{\kappa-}},
\frac{\delta }{\delta \tilde b_{\kappa-}}
 \bigg)} 
\\ 
+Z^{(k)}_C\ensuremath{\bigg(
\frac{\delta }{\delta c_{\kappa+}},
\frac{\delta }{\delta \tilde c_{\kappa+}},
\frac{\delta }{\delta c_{\kappa-}},
\frac{\delta }{\delta \tilde c_{\kappa-}}
 \bigg)}
\bigg]} 
\\ 
= \mathcal{Z}_C\ensuremath{\bigg(
\frac{\delta }{\delta \psi_+},
\frac{\delta }{\delta {\tilde\psi}_+},
\frac{\delta }{\delta \psi_-},
\frac{\delta }{\delta {\tilde\psi}_-}
 \bigg)} . 
\hspace{0.4\columnwidth}\hspace{-0.4\twocolumnwidth} 
\label{eq:42TB} 
\end{multline}}%
The condition, 
{\begin{align}{{
 \begin{aligned} 
& b_{\kappa-}=b_{\kappa+}, 
& & \tilde b_{\kappa-}=\tilde b_{\kappa+}, 
& & c_{\kappa-}=c_{\kappa+}, 
& & \tilde c_{\kappa-}=\tilde c_{\kappa+}, 
\end{aligned}}}%
\label{eq:74XH} 
\end{align}}%
may then be replaced by, 
{\begin{align}{{
 \begin{aligned} 
& \psi_-=\psi_+, & & {\tilde\psi}_-= {\tilde\psi}_+, 
\end{aligned}}}%
\label{eq:75XJ} 
\end{align}}%
as in (\ref{eq:60PP}). We have thus derived the Hori-Wick theorem for the quantum channel from the Hori-Wick theorems for mode operators. 
\section{Extension of the test-case formula (\ref{eq:55TR}) to fermions}%
\label{ch:CF}
Expanding all exponents in the intermediate formula in (\ref{eq:55TR}) in Taylor series we have, 
\begin{widetext} 
{\begin{multline} 
\Phi _{\textrm{vac}}\ensuremath{\big(
\tilde\eta_+,\eta_+,\tilde\eta_-,\eta_-
 \big)}
= \sum _{\ensuremath{\{m,n\}}}
\frac{(-1)^{m+m''+n'+n'''}(i \hbar )^{m+m'+m''+m'''}}
{m!\hspace{1\auxl} m'!\hspace{1\auxl} m''!\hspace{1\auxl} m'''!\hspace{1\auxl} n!\hspace{1\auxl} n'!\hspace{1\auxl} n''!\hspace{1\auxl} n'''!}
\\ \times 
\ensuremath{\bigg(
\varepsilon_{\mathrm{f}} 
\dfrac{\delta }{\delta \psi_+}
\Delta _{\text{F}}
\dfrac{\delta }{\delta \tilde \psi_+}
 \bigg)}^m
\ensuremath{\bigg(
\varepsilon_{\mathrm{f}}
\dfrac{\delta }{\delta \psi_-}
\tilde\Delta _{\text{F}}
\dfrac{\delta }{\delta \tilde \psi_-}
 \bigg)}^{m'}
\ensuremath{\bigg(
\varepsilon_{\mathrm{f}}
\dfrac{\delta }{\delta \psi_-}
\Delta ^{(+)}
\dfrac{\delta }{\delta \tilde \psi_+}
 \bigg)}^{m''}
\ensuremath{\bigg(
\varepsilon_{\mathrm{f}}
\dfrac{\delta }{\delta \psi_+}
\Delta ^{(-)}
\dfrac{\delta }{\delta \tilde \psi_-}
 \bigg)}^{m'''}
\\ \times 
\ensuremath{\big(
\tilde\eta_+\psi_+
 \big)}^{n} 
\ensuremath{\big(
\tilde\eta_-\psi_-
 \big)}^{n'} 
\ensuremath{\big(
{\tilde\psi}_+\eta_+ 
 \big)}^{n''} 
\ensuremath{\big(
{\tilde\psi}_-\eta_-
 \big)}^{n'''}
\settoheight{\auxlv}{$|$}%
\raisebox{-0.3\auxlv}{$|_{\psi_{\pm}=\tilde \psi_{\pm}=0}$} \, . 
\label{eq:59TV} 
\end{multline}}%
We use condensed notation (\ref{eq:60TW}). 
Importantly, all ``building blocks'' in (\ref{eq:59TV}) are bosonic: they contain an even number of anticommuting quantities (either zero or two). Bosonic quantities behave to a large extent as c-number ones, 
{\begin{gather}{{
 \begin{gathered} 
\ensuremath{\big(
\tilde\eta_+\psi_+
 \big)} 
\ensuremath{\big(
\tilde\eta_-\psi_-
 \big)} = 
\ensuremath{\big(
\tilde\eta_-\psi_-
 \big)} 
\ensuremath{\big(
\tilde\eta_+\psi_+
 \big)} , 
\\
\ensuremath{\bigg(
\dfrac{\delta }{\delta \psi_+}
\Delta _{\text{F}}
\dfrac{\delta }{\delta \tilde \psi_+}
 \bigg)}
\ensuremath{\bigg(
\dfrac{\delta }{\delta \psi_-}
\tilde\Delta _{\text{F}}
\dfrac{\delta }{\delta \tilde \psi_-}
 \bigg)} 
= 
\ensuremath{\bigg(
\dfrac{\delta }{\delta \psi_-}
\tilde\Delta _{\text{F}}
\dfrac{\delta }{\delta \tilde \psi_-}
 \bigg)} 
\ensuremath{\bigg(
\dfrac{\delta }{\delta \psi_+}
\Delta _{\text{F}}
\dfrac{\delta }{\delta \tilde \psi_+}
 \bigg)}, 
\end{gathered}}}%
\label{eq:61TX} 
\end{gather}}%
\end{widetext}%
etc. These properties allowed eq.\ (\ref{eq:59TV}) to be derived in the first place. 

Using (\ref{eq:90QW}) it is straightforward to derive the commutational rules, 
{\begin{align}{{
 \begin{aligned} 
\ensuremath{\bigg(
\varepsilon_{\mathrm{f}} 
\dfrac{\delta }{\delta \psi}
K
\dfrac{\delta }{\delta \tilde \psi}
 \bigg)}\ensuremath{\big(
\tilde\eta \psi 
 \big)} & = {
\tilde\eta
K
\dfrac{\delta }{\delta \tilde \psi}
} +\ensuremath{\big(
\tilde\eta \psi 
 \big)}\ensuremath{\bigg(
\varepsilon_{\mathrm{f}} 
\dfrac{\delta }{\delta \psi}
K
\dfrac{\delta }{\delta \tilde \psi}
 \bigg)} , 
\\ 
\ensuremath{\bigg(
\tilde\eta
K
\dfrac{\delta }{\delta \tilde \psi}
 \bigg)}\ensuremath{\big(
\tilde\psi\eta 
 \big)} & = {
\tilde\eta
K
\eta 
}+\ensuremath{\big(
\tilde\psi\eta 
 \big)}\ensuremath{\bigg(
\tilde\eta
K
\dfrac{\delta }{\delta \tilde \psi}
 \bigg)}, 
\end{aligned}}}%
\label{eq:62TY} 
\end{align}}%
where \mbox{$
K(x,x',t-t')
$} is a c-number kernel and $\psi (x,t),{\tilde\psi}(x,t)$ are auxiliary variables, c-number or Grassmann ones. Applying these rules to eq.\ (\ref{eq:59TV}) we can commute all derivatives to the right; any derivative that reaches the rightmost position is set to zero. The result has the structure, 
{\begin{multline}\hspace{0.4\columnwidth}\hspace{-0.4\twocolumnwidth} 
\Phi _{\textrm{vac}}\ensuremath{\big(
\tilde\eta_+,\eta_+,\tilde\eta_-,\eta_-
 \big)}
= \sum_{\ensuremath{\{
m
\}}} C_{mm'm''m'''}\ensuremath{\big(
{\tilde \eta_+}
\Delta _{\text{F}}
{\eta_+}
 \big)}^m
\\ \times 
\ensuremath{\big(
{\tilde\eta_-}
\tilde\Delta _{\text{F}}
{ \eta_-}
 \big)}^{m'}
\ensuremath{\big[
{\tilde\eta_-}
\Delta ^{(+)}
{ \eta_+}
\big]}^{m''}
\ensuremath{\big[
{\tilde\eta_+}
\Delta ^{(-)}
{ \eta_-}
\big]}^{m'''}, 
\hspace{0.4\columnwidth}\hspace{-0.4\twocolumnwidth} 
\label{eq:63TZ} 
\end{multline}}%
where $C_{mm'm''m'''}$ are c-number coefficients. The critical observation is that all fermionic sign factors present in (\ref{eq:59TV}) ``perish'' in commutations (\ref{eq:62TY}). Coefficients $C_{mm'm''m'''}$ therefore coincide for bosons and fermions. In turn, eqs.\ (\ref{eq:75TA}) and (\ref{eq:63TZ}) coincide up to the expansion of $F$ in Taylor series. This suffices in order to extend the test case formula (\ref{eq:55TR}) to fermions. 


\begin{thebibliography}{1}
\bibitem{endMeso}
The term ``mesoscopic'' refers to quantum manifestations under macroscopic conditions. Common as it actually is, this term is confusing. It implies that there exists something ``else'' that distinguishes macroscopic and mesoscopic conditions. In fact, there is nothing ``else'' apart from the hard-working experimental physicist, rescuing fragile quantum effects from being destroyed by environmental and measurement noises.
\bibitem{endRA}
There is little doubt that equations for dressed thermal retarded and advanced Green functions in the Retarded-Advanced formalism will be obtainable in the out-of-equilibrium formalism. However, the latter remains a subject of forthcoming papers. A comprehensive comparison of the two approaches naturally belongs there.
\bibitem{endKern}
Kernels characteristic of the real field in this paper and in \mbox{Ref.\ \cite{API}} are related by the formulae, 
\mbox{$
G^{(+)} = -D
$}, 
\mbox{$
G_{\text{F}} = -D_{\text{F}}
$} and 
\mbox{$
G_{\text{R}} = -D_{\text{R}}
$}.
\bibitem{endPh}
For the complex field, phase conventions in this paper mostly follow \mbox{Ref.\ \cite{APIII}}, except imaginary unit is moved from response substitutions to definitions of characteristic functionals in response picture. To recover formulae of \mbox{Ref.\ \cite{APIII}} is suffices to replace $\eta ,\tilde\eta \to i\eta ,-i\eta ^{\dag}$. Relation between definitions of \mbox{Refs.\ \cite{API,APII}} and \cite{APIII} was discussed in \cite{APIII}, cf.\ eqs.\ (B.12) and (B.23) in appendix B. 

For the real field, differences with \mbox{Refs.\ \cite{API,APII}} are more pronounced. In \cite{API}, external sources were defined with opposite sign, inducing more sign changes. Heed should be payed to the rules, \mbox{$
\eta _{\mathrm{\pm}},J_{\mathrm{e}},\eta ,j_{\mathrm{e}}\to
-\eta _{\mathrm{\pm}},-j,-i\eta ,-\sigma 
$}, cf.\ also endnote \cite{endKern}.
\end{thebibliography}
\end{document}